\documentclass[fleqn,usenatbib]{mnras}

\usepackage{newtxtext,newtxmath}

\usepackage[T1]{fontenc}

\DeclareRobustCommand{\VAN}[3]{#2}
\let\VANthebibliography\thebibliography
\def\thebibliography{\DeclareRobustCommand{\VAN}[3]{##3}\VANthebibliography}

\usepackage{graphicx}	
\usepackage{amsmath}	
\usepackage{amssymb}	
\usepackage{natbib}
\usepackage{hyperref}
\usepackage{caption}
\usepackage{subcaption}
\DeclareGraphicsExtensions{.pdf,.png,.jpg,.mps,.eps,.ps}

\newcommand\nustar{{\it NuSTAR}}
\newcommand\xmm{{\it XMM-Newton}}
\newcommand\s{{\rm~s}}

\newcommand\xiunit{\ifmmode {\rm~erg\s}$^{-1}$ \else ~erg~cm~s$^{-1}$\fi}
\newcommand\kms{\ifmmode {\rm~km\ s}$^{-1}$ \else ~km s$^{-1}$\fi}
\newcommand\Hunit{\ifmmode {\rm~km\ s}$^{-1}$\ {\rm Mpc}$^{-1}$
	\else ~km s$^{-1}$ Mpc$^{-1}$\fi}
\newcommand\cts{\ifmmode {\rm~count\ s}$^{-1}$ \else ~count s$^{-1}$\fi}
\newcommand\ergsec{\ifmmode {\rm~erg\ s}$^{-1}$ \else
	~erg s$^{-1}$\fi}
\newcommand\funit{\ifmmode {\rm~erg\ s}$^{-1}$\;{\rm cm}$^{-2}$ \else
	~erg s$^{-1}$ cm$^{-2}$\fi}
\newcommand\phflux{\ifmmode {\rm~photon\ s}$^{-1}$\;{\rm cm}$^{-i2}$
	\else   ~photon s$^{-1}$ cm$^{-2}$\fi}
\newcommand\efluxA{\ifmmode {\rm~erg\ s}$^{-1}$\;{\rm cm}$^{-2}$\;{\rm
		\AA}$^{-1}$ \else ~erg s$^{-1}$ cm$^{-2}$ \AA$^{-1}$\fi}
\newcommand\efluxHz{\ifmmode {\rm~erg\ s}$^{-1}$\;{\rm cm}$^{-2}$\;{\rm
		Hz}$^{-1}$ \else ~erg s$^{-1}$ cm$^{-2}$ Hz$^{-1}$\fi}
\newcommand\cc{\ifmmode {\rm~cm}$^{-3}$ \else cm$^{-3}$\fi}
\newcommand\cs{\ifmmode {\rm~cm}$^{-2}$ \else cm$^{-2}$\fi}
\newcommand\FWHM{\ifmmode {\rm~FWHM} \else ${\rm~FWHM}$\fi}
\newcommand\Msun{\ifmmode M_{\odot} \else $M_{\odot}$\fi}
\newcommand\Lsun{\ifmmode L_{\odot} \else $L_{\odot}$\fi}

\newcommand\hbeta{\ifmmode {\rm H}\beta \else H$\beta$\fi}
\newcommand\Kalpha{\ifmmode {\rm K}\alpha \else K$\alpha$\fi}
\newcommand\nh{\ifmmode N_{\rm H} \else N$_{\rm H}$\fi}

\title[NGC~4051]{Investigation of a small X-ray flaring event in NLS1 galaxy NGC~4051}

\author[Kumari et al.]{
Neeraj Kumari,$^{1,2}$\thanks{E-mail: neerajkumari@prl.res.in, neerjakumari108@gmail.com}
Arghajit Jana$^{3}$,
Sachindra Naik$^{1}$, Prantik Nandi$^{1}$
\\
$^{1}$ Astronomy and Astrophysics division, Physical Research Laboratory, Ahmedabad-380009, India\\
$^{2}$ Indian Institute of Technology, Gandhinagar-382355, India\\
$^{3}$ Institute of Astronomy, National Tsing Hua University, Hsinchu 30013, Taiwan
}

\date{Accepted XXX. Received YYY; in original form ZZZ}

\pubyear{2021}

\begin{document}
\label{firstpage}
\pagerange{\pageref{firstpage}--\pageref{lastpage}}
\maketitle

\begin{abstract}

We performed a detailed broadband spectral and timing analysis of a small flaring event of $\sim$120 ks in a narrow-line Seyfert 1 galaxy NGC~4051 using simultaneous \xmm~ and \nustar~ observations. The $\sim$300 ks long \nustar~ observation and the overlapping \xmm~ exposure were segregated into pre-flare, flare, and post-flare segments.
During the flare, the \nustar~ count rate peaked at 2.5 times the mean count rate before the flare. Using various physical and phenomenological models, we examined the $0.3-50$ keV X-ray spectrum, which consists of a primary continuum, reprocessed emission, warm absorber and ultra-fast outflows in different timescales. The mass of the central black hole is estimated to be $\ge 1.32\times 10^5 M_{\odot}$ from the spectral analysis. The absence of correlation between flux in 6--7 keV and 10--50 keV bands suggests different origins of the iron emission line and the Compton hump. From the spectral analysis, we found that the reflection fraction drops significantly during the flare, accompanied by the increase in the coronal height above the disc. The spectrum became soft during the flare supporting the "softer when brighter" nature of the source. After the alleviation of the flare, the coronal height drops and the corona heats up. This indicates that there could be inflation of the corona during the flare. We found no significant change in the inner accretion disc or the seed photon temperature. These results suggest that the flaring event occurred due to the change in the coronal properties rather than any notable change in the accretion disc.

\end{abstract}




\section{Introduction}

Active Galactic Nuclei (AGNs) are the accreting super massive black holes (SMBH) of mass $>$ 10$^5$ M$_\odot$, located at the centre of almost every galaxy. These are considered to be the powerful sources (luminosities up to 10$^{48}$\ergsec) in the Universe that emit in the entire range of the electromagnetic spectrum. The ultraviolet/optical photons originating from a thermal accretion disc are inverse Comptonized in the hot corona of relativistic electrons near the black hole and produce X-ray continuum emission \citep{1993ApJ...413..507H}, that can be approximated by a power-law with an exponential cut-off. However, the origin of the compact corona and its nature, structure and location are still unknown. Recent studies put forward the hint of the connection between the corona and the jet \citep{2020MNRAS.496..245Z}. It is hypothesized that in Seyfert galaxies (or radio-quiet AGNs), the corona may be powered by the magnetic flux tubes associated with the accretion disc near the black hole. These flux tubes get inflated due to the differential rotation of the black hole and the accretion disc, and strong confinement from the surrounding medium quenches the launching of jets. As a result, the plasma gets heated up and produces X-ray emission \citep{2019MNRAS.487.4114Y}. The X-ray emission shows high variability in the timing and spectral properties on short as well as long timescales \citep{2021PASA...38...42K}. The rapid and small variability, where the count rate in different energy ranges changes by a factor of two or three, is frequently observed in the AGNs and primarily attributed to the change in the coronal topology. 

Narrow-Line Seyfert galaxies 1 (NLS1s) are the subcategory of AGNs and are believed to be accreting mass at a higher rate or close to the Eddington limit \citep{Komossa_2006}. In the X-ray band, the NLS1s show a steep X-ray spectral slope ($\Gamma \sim$ 1.6 -- 2.5; \citealt{10.1046/j.1365-8711.1999.02811.x}). It is proposed that when the accretion rate is high, more soft photons are produced, causing the corona to cool down rapidly and produce brighter and softer spectra \citep{Shemmer_2008}. The NLS1s are known to show strong reprocessed emission, soft excess below 2 keV and signatures of ionized or neutral absorbers \citep{10.1046/j.1365-8711.2002.05419.x, 1996A&A...309...81W} in their spectra. Moreover, strong outflows have been observed in these objects, which are associated with a high mass accretion rate \citep{2021NatAs...5...13L}. The NLS1s are well known to show strong soft X-ray excess below 2 keV, though the origin of this feature is still debatable. There are different models to explain the origin of this component, e.g., warm Comptonization in relatively cool and optically thick plasma \citep{10.1111/j.1365-2966.2011.19779.x, 2018MNRAS.480.1247K} and relativistic blurred reflection from the inner disc \citep{10.1111/j.1365-2966.2004.08036.x}. Some authors also suggest that fewer scattering in the corona, which can generate higher variability than a large number of scatterings, could also possibly lead to the origin of the soft-excess \citep{2021MNRAS.506.3111N}. 

NGC~4051, a nearby (z=0.00234) galaxy, is one of the brightest NLS1 galaxies with a mass $\sim 1.7\times 10^6$ M$_\odot$ \citep{2009ApJ...702.1353D}. This is a low-luminosity ($\sim $10$^{41}$\ergsec) NLS1 galaxy which shows rapid flux and spectral variability (e.g. \citealt{1996PASJ...48..781G}) on different timescales and a number of emission/absorption features in the spectrum. The X-ray spectral and timing variabilities of this AGN have been studied copiously in the past. Using 1000 days of data from 1996 to 1999, \cite{2003MNRAS.338..323L} reported a very hard spectrum during its low state. There was an underlying reflection emission likely from the molecular torus in all the flux states. They found an anti-correlation between the X-ray flux and spectral hardness. Similar results have been reported  by \cite{10.1111/j.1365-2966.2004.07639.x}, where the reflection of hard X-rays from the cold matter was found along with a non-variable and narrow Iron K$\alpha$ fluorescent line. The difference between the low and high flux spectral states was interpreted in terms of varying properties of the outflowing gas. \cite{2006MNRAS.368..903P} studied the time-resolved spectra using \xmm~ observation and calculated the slope-flux correlation in the low and high flux states. \cite{10.1093/mnras/stw3364} found the presence of different kinds of warm absorbers (WAs) in the X-ray spectrum and their variability. In 2018, \cite{2018A&A...613A..48S} reported a correlation between the X-ray spectral photon-index and the mass accretion rate using the \xmm, {\it Suzaku} and {\it RXTE} observations. This provided an estimate of the lower limit of the black hole mass in NGC~4051 as $6\times 10^5$ M$_\odot$. In this work, we report the results from the X-ray timing and spectral analysis of simultaneous \xmm~ and \nustar~ observations of NGC~4051 in November 2018. We investigated the spectral variability at three time intervals of the observation to examine the changes in the X-ray emitting corona with flux variability. 

This paper is organized as follows: Section~\ref{sec:Observations and Data reduction} describes the observations of NGC~4051 with \xmm~ and \nustar~ observatories and corresponding data reduction processes. The results from the time-resolved spectral analysis and timing analysis are discussed in Section~\ref{sec:Data Analysis and Results}. The implications of the results have been discussed in Section~\ref{sec:Discussion}. In the end, the conclusions have been summarized in Section~\ref{sec:summary}. In this work, we used cosmological parameters; H$_0$=70\Hunit, $\Lambda_0$=70, and $\sigma_M$=0.27.

\section{Observations and Data reduction}
\label{sec:Observations and Data reduction}
NGC~4051 was observed with \nustar~\citep{2013ApJ...770..103H} and \xmm~\citep{2001A&A...365L...1J} multiple times in the last two decades. Most recently, it was observed with \nustar~ in November 2018 simultaneously with \xmm. The observation details are given in Table~\ref{tab:obs}. Two observation IDs of \xmm~ that are used in this work were studied by \cite{2020SCPMA..6329512W}.

\subsection*{\xmm}
In this work, we used the European Photon Imaging Camera (EPIC)-pn \citep{2001A&A...365L..18S} small window mode data due to its higher sensitivity. We reprocessed the EPIC-pn data using standard \textsc{System Analysis Software (SAS v.19.1.0;} \citealt{2004ASPC..314..759G}) and updated calibration files. To check the presence of any flaring particle background, we generated a light curve above 10 keV. We created good time interval (GTI) files by excluding the time intervals containing particle background, which were capped by the count rate. We used $<$0.15 and $<$0.2 \cts~ for 0830430801 and 0830430201 Obs. IDs, respectively. This resulted in corresponding net exposures of $\sim73.8$ and $\sim68.8$ ks. We considered events with {\tt PATTERN}$\le$4 for EPIC-pn. We also checked for the photon pile-up effect using the {\tt epatplot} task. We extracted source and background spectra by selecting circular regions of 30 arcsec radii for both the observations. We generated light curves in different energy bands using the {\tt xselect} task of {\tt FTOOLS}. The redistribution matrix and ancillary response files were generated by {\tt arfgen} and {\tt rmfgen} tasks, respectively. We grouped the spectral data sets to a minimum of 30 counts per bin using the {\tt grppha} task.

\subsection*{\nustar}
\nustar~ consists of two hard X-ray focusing telescopes called Focal Plane Modules FPMA and FPMB. The level~1 data files from both the modules were reprocessed with the \textsc{\nustar~ Data Analysis Software (NuSTARDAS v2.0.0)} package incorporated in the latest {\tt heasoft} package (v6.28). The level~2 cleaned and calibrated data products were produced using {\tt nupipiline} task with the latest calibration files (CALDB)\footnote{\url{http://heasarc.gsfc.nasa.gov/FTP/caldb/data/nustar/fpm/}}. We considered circular regions of 80 arcsec radii to extract source and background spectra from the event files. The source region was centered at the coordinates of the optical counterpart, and the  background region was selected away from the source. The {\tt nuproducts} task was used to generate the spectra and light curves. We generated time-resolved spectra (pre-flare, flare and post-flare) by using the {\tt nuproduct} task. We also generated spectra for the rising and decay phases of the flare separately. The spectra are binned with 30 counts per bin. We used the spectral range up to 50 keV where the count rate was significantly higher than the background. 

\begin{table*}
	\centering
		\caption{Log of November 2018 simultaneous observations of NGC~4051 with \nustar~and \xmm. The quoted count rate for \nustar~ observation is the average count rate over the entire observation.} 
		\label{tab:obs}
		\begin{tabular}{lccccccc}
			\hline
			UT Date (start) & UT Date (end) & \nustar & Total Exp (ks) & Counts s$^{-1}$ & \xmm & Exp (ks) & Counts s$^{-1}$ \\
			\hline 
			2018-11-04 12:56:09 & 2018-11-11 14:46:09 & 60401009002 & 311.1 & 0.45 & -- & -- & -- \\
			2018-11-07 09:59:39 & 2018-11-08 10:26:15 & -- & -- & -- & 0830430201 & 83.2 & 13.84 \\
			2018-11-09 09:48:58 & 2018-11-10 09:02:59 & -- & -- & -- & 0830430801 & 85.5 & 8.59 \\ 
			\hline
		\end{tabular}
\end{table*}
 
\begin{figure} 
	\centering
	\includegraphics[scale=0.45, angle=0 ,trim={0.3in 0.45in 0 0}, clip]{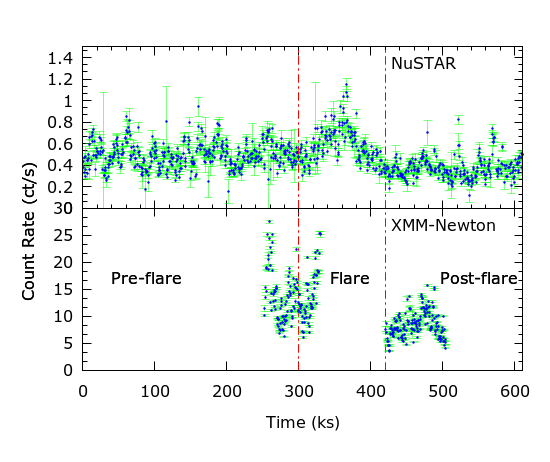} 
	\caption{The X-ray light curves of NGC~4051 in $3-50$ keV and $0.3-10$ keV ranges, obtained from the November 2018 \nustar~(upper panel) and \xmm~(bottom panel) observations, respectively, are shown for 500~s time bin. In the \nustar~ light curve, a flare of $\sim$120~ks duration with peak count rate reaching by a factor of $\sim$2.5 compared to the quiescent duration is detected. The red dotted lines divide the light curves (observations) into three time intervals such as pre-flare, flare and post-flare segments.}
	\label{fig:obs_net1}
\end{figure}

\begin{figure}
    \centering
    \includegraphics[scale=0.43]{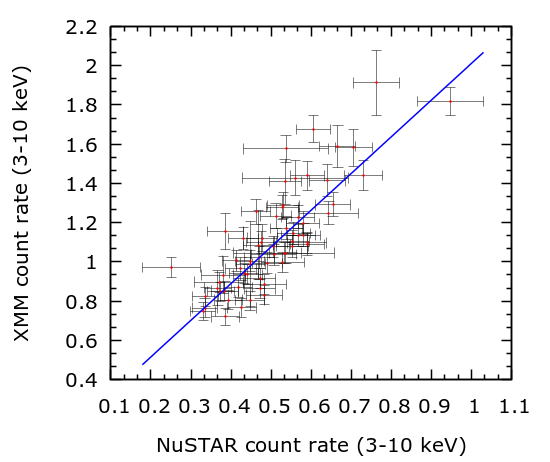}
    \caption{The count rates in 3-10 keV range are shown for the overlapping pre-flare phase of the \xmm~and \nustar~observations.}
    \label{fig:3-10_nustar_xmm}
\end{figure}

 In Figure~\ref{fig:obs_net1}, the light curves of the source from the \nustar~and \xmm~observations are shown. The \nustar~light curve showed the occurrence of a small flare at around $3\times10^5$~s from the beginning of the observation that lasted for $\sim120$~ks (duration between two vertical lines in Figure~\ref{fig:obs_net1}). During this flare, the peak count rate was about 2.5 times the average count rate observed before and after the flare. This flaring event was perhaps prominent in the \xmm~light curve, though the lack of observation coverage during the flare did not make it evident. However, to avoid ambiguity, we tried to correlate the $3-10$ keV count rate for overlapping pre-flare phase from the \xmm~and \nustar~observations (see Figure~\ref{fig:3-10_nustar_xmm}) and found Pearson's correlation coefficient to be $\sim0.82$ with p-value $<$0.01.

\section{Data Analysis and Results}
 \label{sec:Data Analysis and Results}
\subsection{Timing analysis}

\subsubsection*{Fractional variability}
\begin{figure} 
\centering
\includegraphics[scale=0.3, angle=0 ,trim={0 0 0 0}, clip]{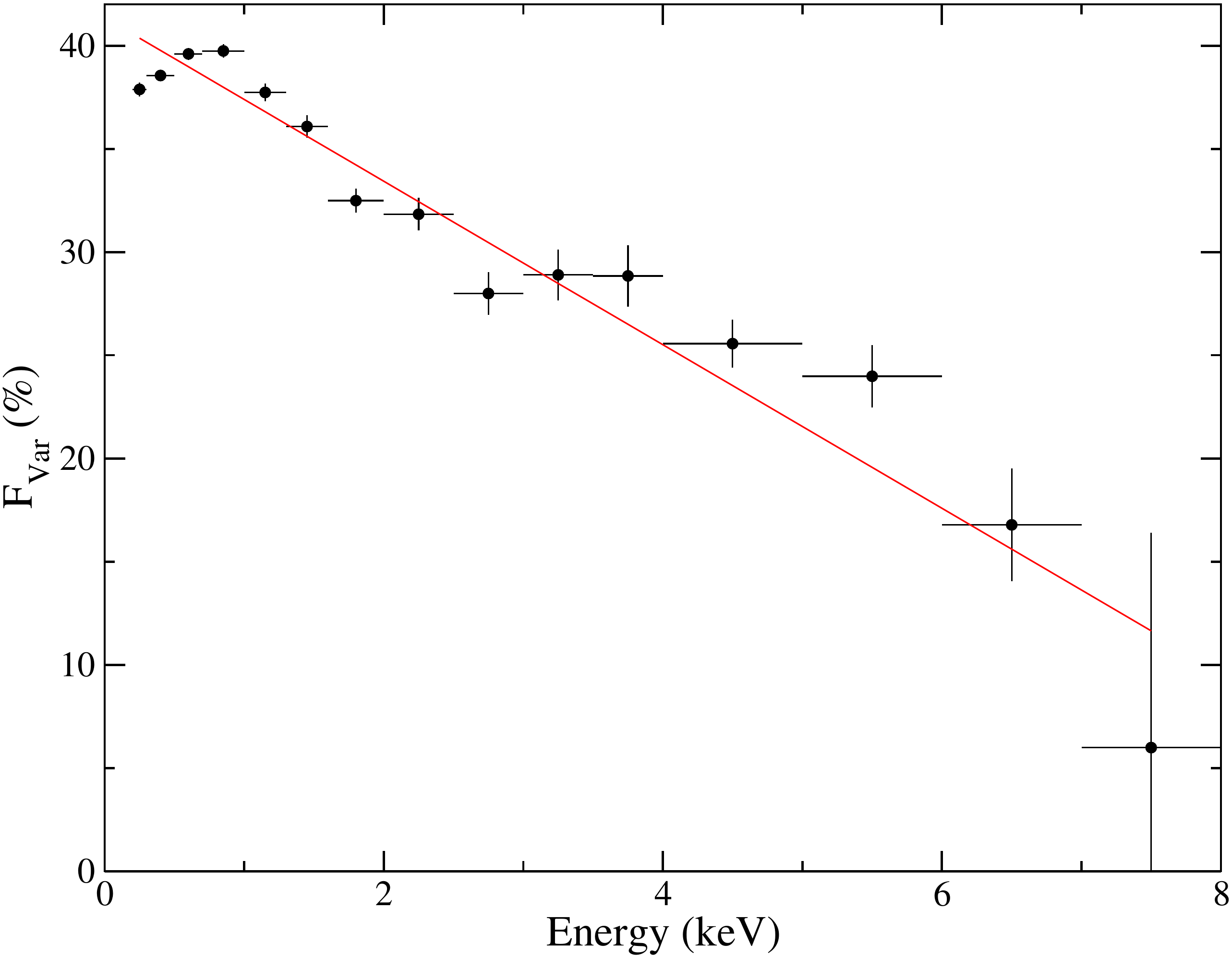} 
\caption{The fractional variability spectrum is shown for the EPIC-pn data from the \xmm~ observation (Obs. ID 083043201) of NGC~4051. The red line represents the linear fit to the observed data points.}
\label{fvar} 
\end{figure}

In the beginning, we attempted to investigate the timing variability of NGC~4051 using \xmm~ and \nustar~ observations. To check the temporal variability, we calculated the fractional variability ($F_{\rm var}$; \citealt{Nandra_1997}, \citealt{10.1046/j.1365-2966.2003.07042.x}) in different energy bands using the equation, 
 
$$F_{\rm var}=\sqrt{\frac{S^2-\overline{\sigma^2_{\rm err}}}{\Bar{x}^2}}$$ 
 
and uncertainty in $F_{\rm var}$ as, 
 
$$\sqrt{\Bigg( \sqrt{\frac{1}{2N}} \frac{\overline{\sigma^2_{\rm err}}}{\Bar{x}^2 F_{\rm var}} \Bigg)^2 + \Bigg( \sqrt{\frac{\overline{\sigma^2_{\rm err}}}{N}} \frac{1}{\Bar{x}}\Bigg)^2 }$$
 
where $\Bar{x}$, $S$, $\overline{\sigma}_{\rm err}$, and $N$ are the mean, total variance, mean error and number of data points, respectively. We generated the variability spectrum, i.e. $F_{\rm var}$ as a function of energy (see Figure~\ref{fvar}) using background subtracted EPIC-pn (obs.~ID 0830430201) light curves in increasing energy binning. Pearson's correlation coefficient was calculated to be $-0.95$ with the null hypothesis probability of $\sim10^{-9}$. The spectrum shows that the low energy components, e.g., soft X-ray excess or/and ionized absorbers are more variable than the high energy ones. 

\begin{figure} 
\centering
\includegraphics[scale=0.49, angle=0 ,trim={0.3in 0 0 0}, clip]{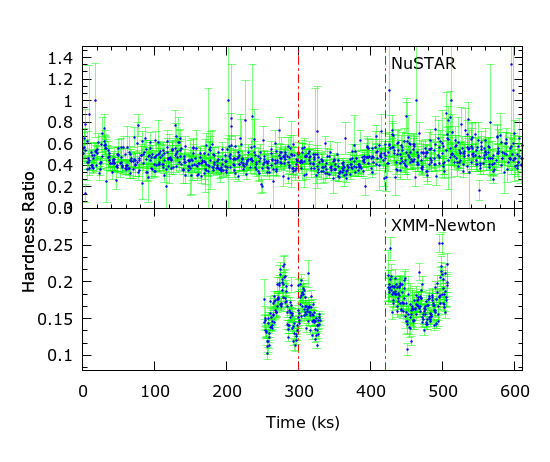} 
\caption{The hardness ratios are shown for \nustar~ (upper panel) and \xmm~ (bottom panel) observations of NGC~4051. The hardness ratios are obtained by taking ratio between the light curves in 10-50 keV and 3-10 keV bands (for \nustar), and 2-10 keV and 0.3-2 keV bands (for \xmm).}
\label{fig:HR} 
\end{figure}

\begin{figure*} 
\centering
\includegraphics[scale=0.45, angle=0 ,trim={0 0 0 0}, clip]{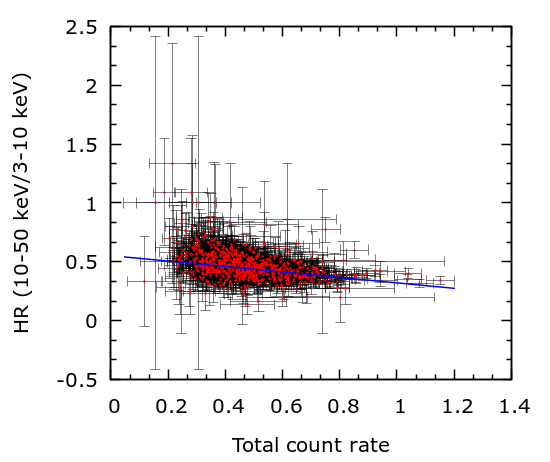} 
\includegraphics[scale=0.45, angle=0 ,trim={0 0 0 0}, clip]{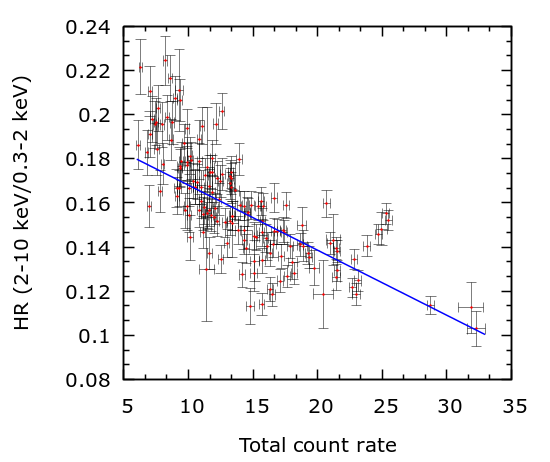} 
\caption{Hardness-Intensity diagrams of NGC~4051 are shown from the \nustar~ (left panel) and \xmm~ (right panel) observations. Corresponding correlation coefficients are $-0.42$ and $-0.73$, respectively.}
\label{fig:HRvsflux} 
\end{figure*}

To have an initial guess of the spectral variability during the observation, we plotted  variation of the hardness ratio (HR), i.e. the ratio between the hard X-ray and soft X-ray light curves, with respect to time. For \nustar~ observation, we computed HR by taking the ratio between the light curves in the $10-50$ keV and $3-10$ keV energy ranges, whereas, for \xmm~ observation, we estimated HR by taking the ratio between the $2-10$ keV and $0.3-2$ keV light curves. From Figure~\ref{fig:HR}, we observed considerable spectral variability in the soft X-rays. The spectrum became soft at the beginning of the flare. Having observed closely, we find the `softer when brighter' trend, which is corroborated by the hardness-intensity diagram (HR vs total count rate) in Figure~\ref{fig:HRvsflux}. On the other hand, the variability seen in the \nustar~ spectrum is comparatively less. The \nustar~ HR is more or less constant except softening marginally during the flaring phase (see the upper panel of Figure~\ref{fig:HR} and left panel of Figure~\ref{fig:HRvsflux}).

\subsubsection*{Correlation}

\begin{figure*}
		\centering
		\includegraphics[scale=0.09]{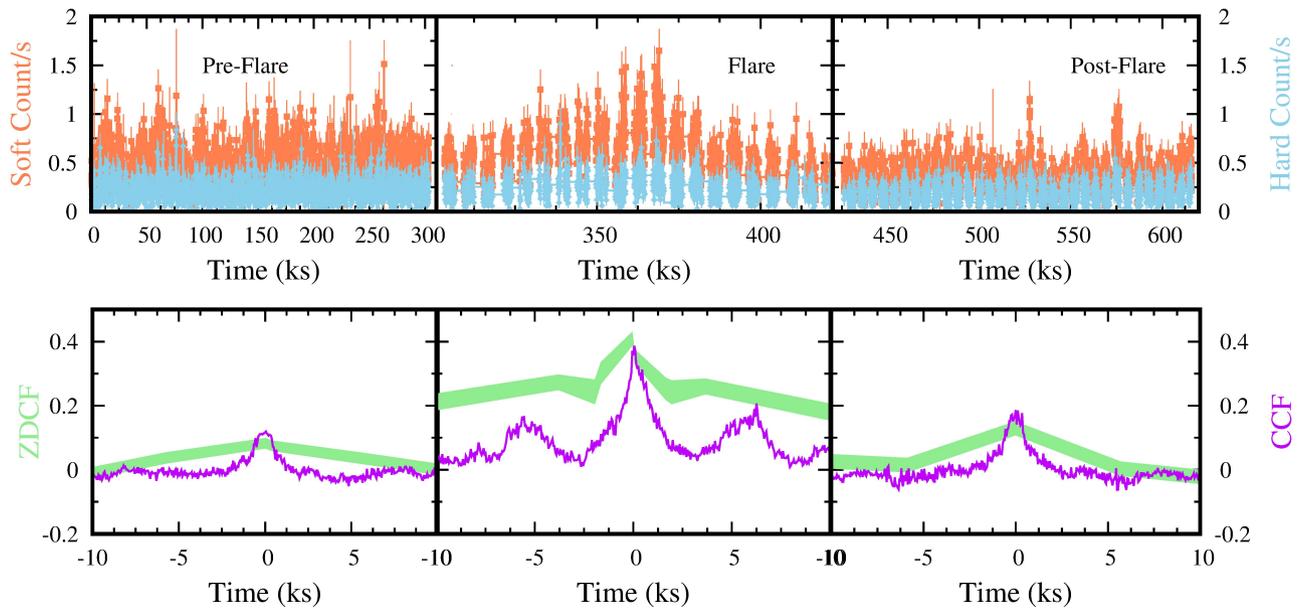}
		\caption{Top panel: The lightcurves of NGC~4051 in soft X-ray (3.0 -- 10.0 keV: salmon) and hard X-ray (10.0 -- 50.0 keV: light blue) bands obtained from {\it NuSTAR} observation are shown for pre-flare, flare and post-flare segments. Bottom panel: The discrete cross-correlations between the soft X-ray and hard X-ray light curves are plotted. A moderately strong correlation between these bands can be seen during the duration of the flare, which is absent in the pre-and post-flare segments. The Cross-Correlation Functions (CCFs) are presented in a solid-magenta line, whereas the green bands represent the $\zeta$-discrete cross-correlations.}
		\label{fig:corr}
	\end{figure*}

For correlation analysis in the X-ray band ($3.0-50.0$ keV), we divided the entire \nustar~observation of NGC~4051 into the pre-flare (MJD 58426.5 to MJD 58430.0), flare (MJD 58430.0 to MJD 58431.4) and post-flare (MJD 58431.4 to MJD 58433.6) segments that are marked with vertical lines in Figure~\ref{fig:obs_net1}. For each segment, we segregated the X-ray light curve into soft X-ray (3.0 to 10.0 keV) and hard X-ray (10.0 to 50.0 keV) bands. Corresponding light curves with time resolutions of 50~s are shown in the top panels of Figure~\ref{fig:corr}. During the entire duration of the observation, the source count rate in the hard X-ray band was always less than that of the soft X-ray band. At the peak of the flaring event, the source count rate in both bands increased by a factor of 2-3 compared to the average count rate during the pre-and post-flare segments. We performed  cross-correlation analysis using {\tt crosscor}\footnote{\url{https://heasarc.gsfc.nasa.gov/xanadu/xronos/help/crosscor.html}} and $\zeta$-discrete cross-correlation function ({\tt ZDCF}\footnote{ \url{https://www.weizmann.ac.il/particle/tal/research-activities/software}}; \citealt{1997ASSL..218..163A}) for comparison of light curves in two different bands. For the error estimation, we considered 12000 simulation points in the {\tt ZDCF} code for the light curves. The soft and hard band light curves during the pre-and post-flare segments yielded an acceptable $\rm \chi^2_{red} < 1.5$ when fitted with a straight line. However, the data during the flare yielded poor statistics with high residuals when fitted with a linear function. We carried out delay estimation through the discrete correlation function using {\it NuSTAR} data during three segments separately.
	
The discrete correlation function analysis, performed on the light curves from three different segments, generated three different patterns as shown in the bottom panels of Figure~\ref{fig:corr}. In the pre-flare and post-flare phases, the soft and hard X-ray bands are found to be very poorly correlated. The values of the correlation function for pre-flare and post-flare segments were found to be $<$0.1 (left bottom panel of Figure~\ref{fig:corr}) and $<$0.2 (right bottom panel of Figure~\ref{fig:corr}), respectively, for the time delay range of $-10$ to $+10$ kilo-second (ks). However, the flaring event data showed a moderately strong correlation between the soft and hard bands in both the algorithms. From the ZDCF, we found the peak of the correlation function at $0.0\pm0.05$ ks with a correlation coefficient of $0.48\pm0.02$, and for the CCF, calculated through {\tt crosscor}, the peak appears at $0.0\pm0.05$ with a correlation coefficient of $0.4$ (middle bottom panel of Figure~\ref{fig:corr}). Although from the CCF analysis of post-flare segment data, a minor peak is observed near $0.0\pm50$ ks time delay, it is not prominent as  the peak value is $0.13$ (bottom right panel of Figure~\ref{fig:corr}). Therefore, we are not considering this as a peak in the correlation function. We do not notice any significant peak in the $\zeta$-discrete cross-correlation function in this post-flaring segment. From the correlation analysis, we conclude that the soft and hard X-ray light curves are moderately correlated only during the flare segment.

\subsection{Spectral analysis}
We used {\tt XSPEC} (v12.11.1) software package \citep{1996ASPC..101...17A} to analyse the spectral data and $\chi^2$ statistics to find the best-fitting models. Throughout the spectral analysis, the uncertainties in various parameters are calculated using {\tt `error'} command in {\tt XSPEC} at 90\% confidence level unless or otherwise stated. As mentioned earlier, the data from the \nustar~ and \xmm~ observations are divided into three segments such as pre-flare, flare and post-flare segments (see Figure~\ref{fig:obs_net1}). We carried out the spectral analysis in the $0.3-50$ keV energy range using simultaneous \xmm~ and \nustar~ observations for all three segments.   

The X-ray spectrum of an AGN is typically described by a power-law continuum, complex iron emission lines in $\sim 6-7$ keV range, reflection hump in $\sim 20-40$ keV range and soft X-ray excess below 2 keV. We used various phenomenological as well as physical models and included all the features in our analysis. We used {\tt zpowlw}\footnote{\url{https://heasarc.gsfc.nasa.gov/xanadu/xspec/manual/node212.html}}, {\tt borus} \citep{2018ApJ...854...42B}, and {\tt relxilllpion}, modified version of {\tt relxill}\footnote{\url{http://www.sternwarte.uni-erlangen.de/~dauser/research/relxill/}} model \citep{2016A&A...590A..76D}. For Galactic absorption and local ionized absorption, we used {\tt phabs} or {\tt tbabs} \citep{2000ApJ...542..914W} and {\tt zxipcf}\footnote{\url{https://heasarc.gsfc.nasa.gov/xanadu/xspec/models/zxipcf.html}} models, respectively.  

We started our spectral analysis with \nustar~ observation in the $3-50$ keV range to have an initial idea of spectral changes with the flux variability. As mentioned earlier, we dissected \nustar~ observation into three parts: pre-flare, flare and post-flare segments. To investigate the spectral changes during the flare, we further divided the flare duration into two sub-segments, such as the rising phase and the declining phase. Our basic model reads in {\tt XSPEC} as: 
\begin{multline*}
\tt phabs\times(atable\{borus\}+zphabs\times cabs \times zcutoffpl \\
\tt + constant\times zcutoffpl)
\end{multline*}

In the above expression, we used the cutoff power-law model to fit the primary continuum. In this model, `{\tt constant}' stands for the relative normalization of leaked or scattered unabsorbed intrinsic continuum. The {\tt zphabs $\times$ cabs} component represents the line-of-sight absorption, including the Compton scattering losses out of the line of sight. Here, the additive {\tt borus} model is used for the reprocessed emission from the cold and neutral gas. This model calculates the fluorescent line emission and reprocessed continuum self-consistently. We simultaneously fitted all four time-resolved spectra with the above model. While fitting, the inclination angle, covering factor ($\rm C_{F, Tor}$) and average column density ($\rm N_{H,Tor}$) of the obscuring materials are tied across the four segments. These parameters are unlikely to change during such a short observation span. The foreground Galactic absorption column density is fixed at $N_{\rm H, Gal}=1.2\times10^{20}$\cs \citep{2016A&A...594A.116H} in the source direction and modelled with {\tt phabs}. We obtained a good fit ($\chi^2$/dof=1669/1603) with this model. All the fitted parameters are listed in Table~\ref{tab:borus}. From this fitting, we found that the photon index ($\Gamma$) was relatively higher during the flare phase as compared to pre- and post-flare phases. We could not constrain the cutoff energy (E$_{\rm cut}$) in all four phases. The inclination angle obtained from this fitting was found to be around 30 degrees which is reasonable for type-1 Seyfert galaxies.


\begin{figure} 
\centering
\includegraphics[scale=0.9, angle=0 ,trim={0.25in 0 0 1.0}, clip]{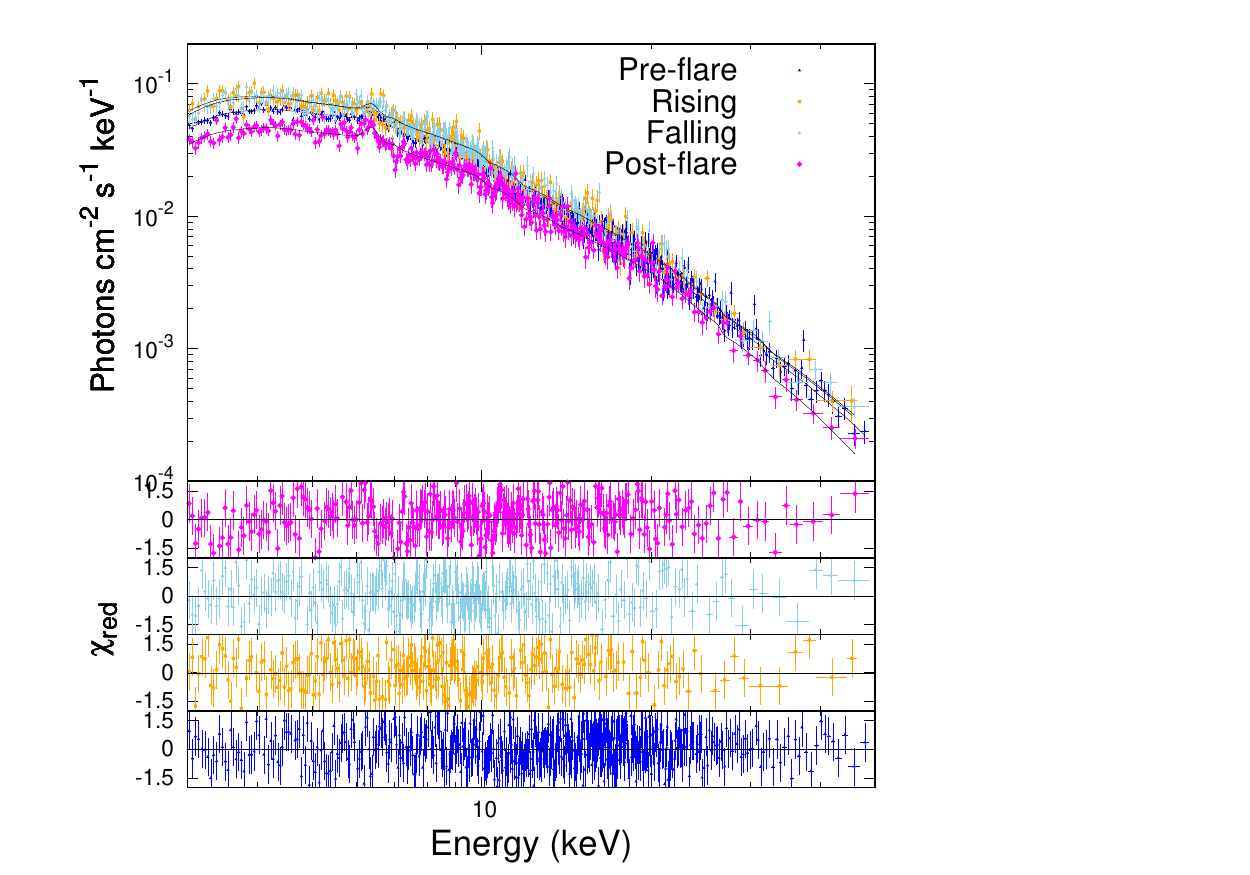} 
\caption{Top panel: The best fit model spectra (fitted simultaneously) for the pre-flare, rising and falling part of the flare, post-flare phases in the $3-50$ keV energy range from \nustar~ observation. Bottom four panels show corresponding residuals. The magenta, cyan, orange and green points represent the data from the pre-flare, rising part of the flare, declining part of the flare and post-flare phases, respectively.}
\label{borus} 
\end{figure}

\begin{table}
	\caption{The best-fitted parameters obtained by fitting the $3-50$ keV \nustar~ time-resolved spectra of NGC~4051 with the {\tt borus} model. }
	\label{tab:borus}
	\begin{tabular}{lcccc}
		\hline
		Parameters & Pre-flare &Rising &Declining & Post-flare \\
		           &segment  &phase  &phase &segment\\
		\hline
		$N_{\rm H, \rm Tor}$ ($10^{22}$ \cs) & $24.50_{-0.06}^{+0.07}$ & $\rm f$& $\rm f$ & $\rm f$ \\
		\\
		$C_{\rm F, \rm Tor}$ & $0.87_{-0.03}^{+0.07}$ & $\rm f$ & $\rm f$ & $\rm f$ \\
		\\
		Incl (deg.) & $30_{-5}^{+5}$ & $\rm f$ & $\rm f$& $\rm f$ \\
		\\
		$A_{\rm Fe}$ & $1.10_{-0.14}^{+0.33}$ & $0.97_{-0.23}^{+0.27}$ & $1.01_{-0.28}^{+0.35}$ & $0.82_{-0.17}^{+0.22}$ \\
		\\
		Norm ($10^{-3}$) & $6.20_{-1.10}^{+1.10}$ & $7.42_{-0.92}^{+0.32}$ & $8.09_{-1.00}^{+0.66}$ & $3.29_{-0.73}^{+1.30}$ \\
		\\
		$N_{\rm H, \rm ref}$ ($10^{22}$ \cs) & $1.22_{-0.55}^{+0.50}$ & $<0.62$ & $0.83_{-0.70}^{+0.59}$ & $1^{\dagger}$\\
		\\
		$\Gamma$ & $1.89_{-0.08}^{+0.06}$ & $1.97_{-0.09}^{+0.05}$ & $1.97_{-0.08}^{+0.03}$ & $1.64_{-0.07}^{+0.09}$ \\
		\\
		\hline
	\end{tabular}
	\leftline{$\rm f$ represents the parameter fixed at the corresponding value at the pre-flare phase.}
	\leftline{$\dagger$ pegged at this value} 
\end{table}

It was found that the photon index remained unchanged during the rising and declining phases of the flare. Therefore, we clubbed these two time intervals together for further spectral analysis. Next, we fitted simultaneous \xmm~ ($0.3-10$ keV range) and \nustar~($3-50$ keV range) spectra in three segments (pre-flare, flare and post-flare) together using different models. The advantage of fitting spectra of three intervals simultaneously is that it produces spectral variability in a more physical way. 

{\bf Model-1:} We built our baseline model with the power law with a high energy cutoff as the source continuum. We used a redshifted black body component ({\tt zbbody}) for the soft excess, a Gaussian line ({\tt zgaus}) for the Fe K-$\alpha$ emission line and the {\tt pexrav} model for the reflection component above 10 keV \citep{1995MNRAS.273..837M}. We set the reflection fraction ($R_{\rm refl}$) as negative so that the {\tt pexrav} component is considered as only the reflection spectrum. We fixed the inclination angle at 30 degrees as obtained from the previous model. The iron and heavy element abundances were tied for three intervals. The photon index of the {\tt pexrav} model was linked with that of the cutoff power-law model. Since the cut-off energy could not be constrained, we fixed it at 370 keV for the low Eddington Seyfert galaxies as suggested in the statistical study by \cite{2017ApJS..233...17R}. We needed two absorbers while fitting the spectra to get the best-fit model. The final model in {\tt XSPEC} reads as,

\begin{multline*}
	\tt constant\times tbabs \times zxipcf \times zxipcf \times (zbbody+ \\
	\tt pexrav+zcutoffpl)
\end{multline*}

We applied a constant factor in the composite model to take into account the calibration uncertainties between different instruments. This model provided a good fit to the \xmm~ and \nustar~ data with $\chi^2=5385$ for 5104 degrees of freedom (dof). We required two ionized absorbers for the residuals below 2 keV. To take into account this absorption feature, we included two {\tt zxipcf} models for partially covering and partially ionized absorbing material. The obtained redshift ($z\sim -0.20$) of the first absorber was similar to the ultra-fast outflows (UFOs)  \citep{10.1093/mnras/sts692} during the three intervals. The second absorber showed a redshift comparable to the warm absorbers (WAs) \citep{2005A&A...431..111B}. The value of the blackbody temperature was found to be $\sim$100 eV which was non-variable during the observation period. The spectrum became softer during the flare with the best-fitted photon-index $\Gamma=2.10\pm0.01$. As we are considering only the reflection component of the {\tt pexrav} model and using a separate continuum model, we could not constrain the value of the reflection scaling factor and fixed it at -1. We calculated the soft excess and power-law flux (see Table~\ref{tab:lumin}) from the corresponding model components. The soft excess flux was found to be decreasing with time while the power-law flux was highest during the flaring phase.  

\begin{table}
	\caption{The best fitted parameters obtained from fitting 0.3-50 keV time-resolved spectra simultaneously using \xmm~ and \nustar~ observations with phenomenological model {\tt zcutoffpl}.}
	\begin{tabular}{lccc}
		\hline
		Parameters & Pre-flare & Flare & Post-flare \\
		           &segment &segment &segment\\
		\hline
		$N_{\rm H,1}$ ($10^{22}$ \cs) & $9.38_{-1.61}^{+1.22}$ & $14.87_{-2.86}^{+2.47}$ & $11.13_{-1.69}^{+1.65}$ \\
		\\
		$\log {\rm \xi_{\rm 1}}$ & $1.08_{-0.95}^{+0.08}$ & $1.09_{-0.56}^{+0.14}$ & $1.31_{-0.09}^{+0.06}$ \\
		\\ 
		Cov Frac1 & $0.24_{-0.01}^{+0.02}$ & $0.20_{-0.01}^{+0.02}$ & $0.23_{-0.03}^{+0.01}$ \\
		\\
		z1 & $-0.20_{-0.06}^{+0.04}$ & $-0.21_{-0.07}^{+0.04}$ & $-0.22_{-0.05}^{+0.02}$  \\
		\\
		\hline
		$\nu$/c & -0.20 & -0.21 & -0.22 \\
            r$_{\rm max}$ (pc) & 0.17 & 0.11 & 0.06 \\
		r$_{\rm min}$ ($10^{-4}$pc) & 0.04 & 0.04 & 0.03 \\
		r$_{\rm min}$ ($R_s$) & 25 & 23 & 21 \\
		\hline
		$N_{\rm H,2}$ ($10^{22}$ \cs) & $0.62_{-0.51}^{+0.52}$ & $2.28_{-0.89}^{+0.79}$ & $2.99_{-0.55}^{+0.52}$ \\
		\\
		$\log {\rm \xi_{\rm 2}}$ & $2.07_{-0.45}^{+0.15}$ & $2.08_{-0.14}^{+0.14}$ &$2.01_{-0.03}^{+0.07}$ \\
		\\
		Cov Frac1 & $0.39_{-0.09}^{+0.58}$ & $0.30_{-0.05}^{+0.07}$ & $0.40_{-0.02}^{+0.03}$ \\
		\\
		z2 & $0.04_{-0.02}^{+0.02}$ & $0.02_{-0.02}^{+0.02}$ & $0.03_{-0.01}^{+0.01}$ \\
		\\
		\hline
            $\nu$/c & 0.04 & 0.02 & 0.03 \\
		r$_{\rm max}$ (pc) & 0.26 & 0.07 & 0.04 \\
		r$_{\rm min}$ ($10^{-4}$pc) & 1.20 & 0.40 & 1.08 \\
		$_{\rm min}$ ($R_s$) & 730 & 3460 & 1372 \\
		\hline

		$kT_{\rm BB}$ (eV) & $100_{-1}^{+1}$ & $100_{-2}^{+2}$ & $100_{-1}^{+1}$ \\ 
		\\
		Norm$_{\rm BB}$ (10$^{-04}$) & $1.50_{-0.02}^{+0.02}$ & $1.11_{-0.22}^{+0.25}$ & $0.85_{-0.01}^{+0.01}$ \\
		\\
		Fe K$\alpha$ ~LE (keV) & $6.40_{-0.03}^{+0.02}$ & $6.36_{-0.05}^{+0.04}$ & $6.35_{-0.02}^{+0.02}$ \\
		\\

		EW (eV) & $101_{-15}^{+16}$ & $82_{-20}^{+22}$ & $95_{-32}^{+29}$ \\
		\\
		Norm ($10^{-5}$) & $1.97_{-0.29}^{+0.32}$ & $1.58_{-0.38}^{+0.42}$ & $1.35_{-0.23}^{+0.23}$ \\
		\\
		Abund & $1.87_{-0.16}^{+0.18}$ & $\rm f$ & $\rm f$ \\
		\\
		Abund$_{\rm Fe}$ & $2.39_{-0.21}^{+0.22}$ & $\rm f$  & $\rm f$ \\

		\\
		$\Gamma$ & $2.05_{-0.01}^{+0.01}$ & $2.10_{-0.01}^{+0.01}$ & $2.05_{-0.01}^{+0.01}$ \\
		\\
		Norm$_{\rm PL}$ (10$^{-03}$) & $7.82_{-0.02}^{+0.02}$ & $8.34_{-0.03}^{+0.03}$ & $5.43_{-0.01}^{+0.01}$ \\
  	    \\
       \hline
	  
	\end{tabular}
	\leftline{$\rm f$ represents the parameter fixed at the corresponding value at the pre-flare phase.}
	\label{tab:zcutof}
\end{table}

{\bf Model-2:} Time-resolved spectral fitting of \nustar~ and \xmm~ data with Model-1 provided information on the spectral changes, presence of warm absorber and UFO in the three time intervals. However, it did not provide any physical properties of the Comptonizing plasma, i.e., the corona. To get information on the change in coronal properties, we replaced {\tt zcutoffpl} with the Comptonization model {\tt compPS} 
\citep{1996ApJ...470..249P}. This model produces X-ray continuum for different geometries using the exact numerical solution of the radiative transfer equation, which depends on the geometry, optical depth of the hot electron plasma, spectral distribution of the seed photons, inclination angle and the way seed soft photons are injected into the hot plasma. This model also takes into account the reflection spectra from the cold medium as well as blurred reflection smeared out by the rotation of the disc and relativistic effects. In the fitting process, we fixed certain parameters based on a prior guess of the system so that fitting could converge. The composite model in {\tt XSPEC} reads as;
\begin{multline*}
	\tt constant\times tbabs \times zxipcf \times zxipcf \times (zgauss+ compPS)
\end{multline*}

\begin{figure*}
     \centering
     \hspace{-0.5in}
     \begin{subfigure}[b]{0.3\textwidth}
         \centering
         \includegraphics[width=1.7\textwidth]{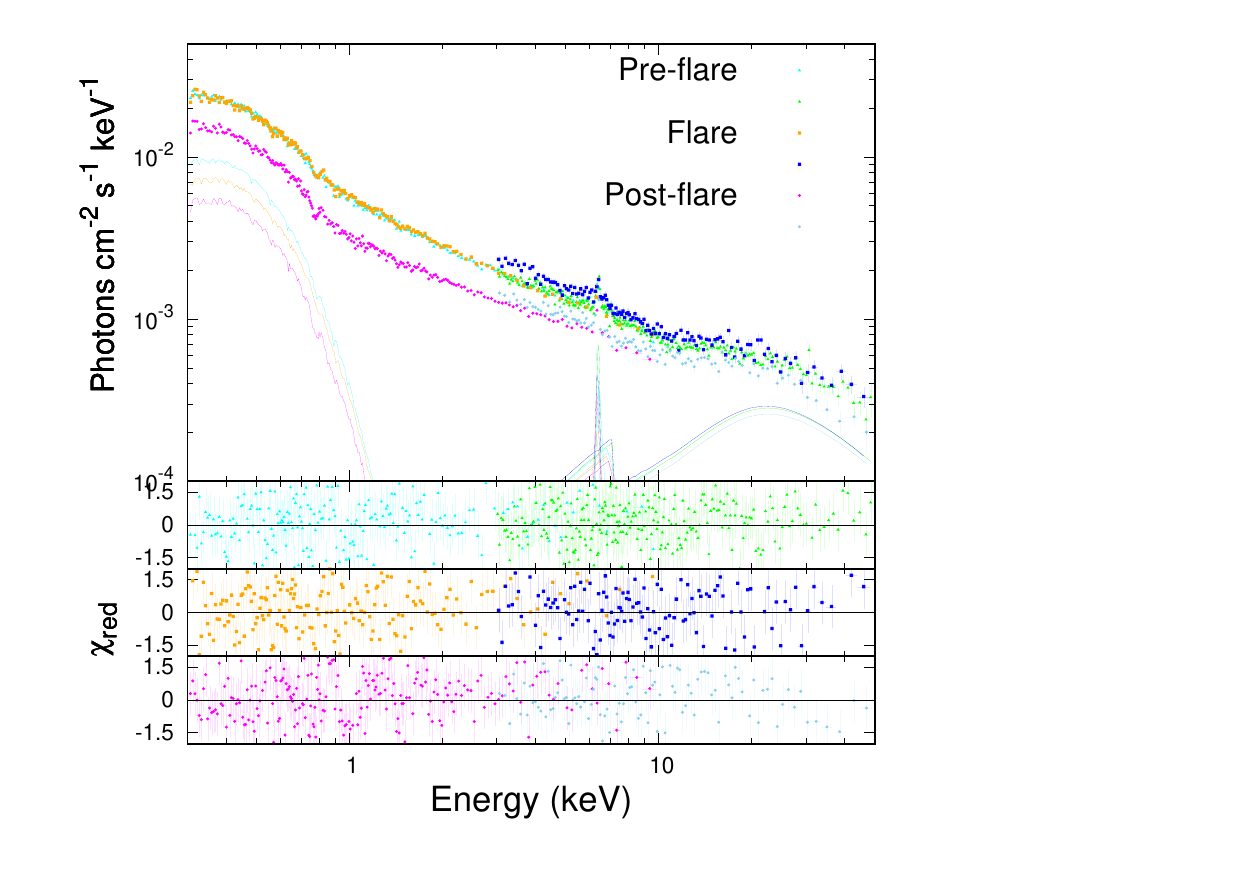}
         \caption{Zcutoff}
         \label{fig:zcutoff}
     \end{subfigure}
     \hspace{0.2in}
     \begin{subfigure}[b]{0.3\textwidth}
         \centering
         \includegraphics[width=1.7\textwidth]{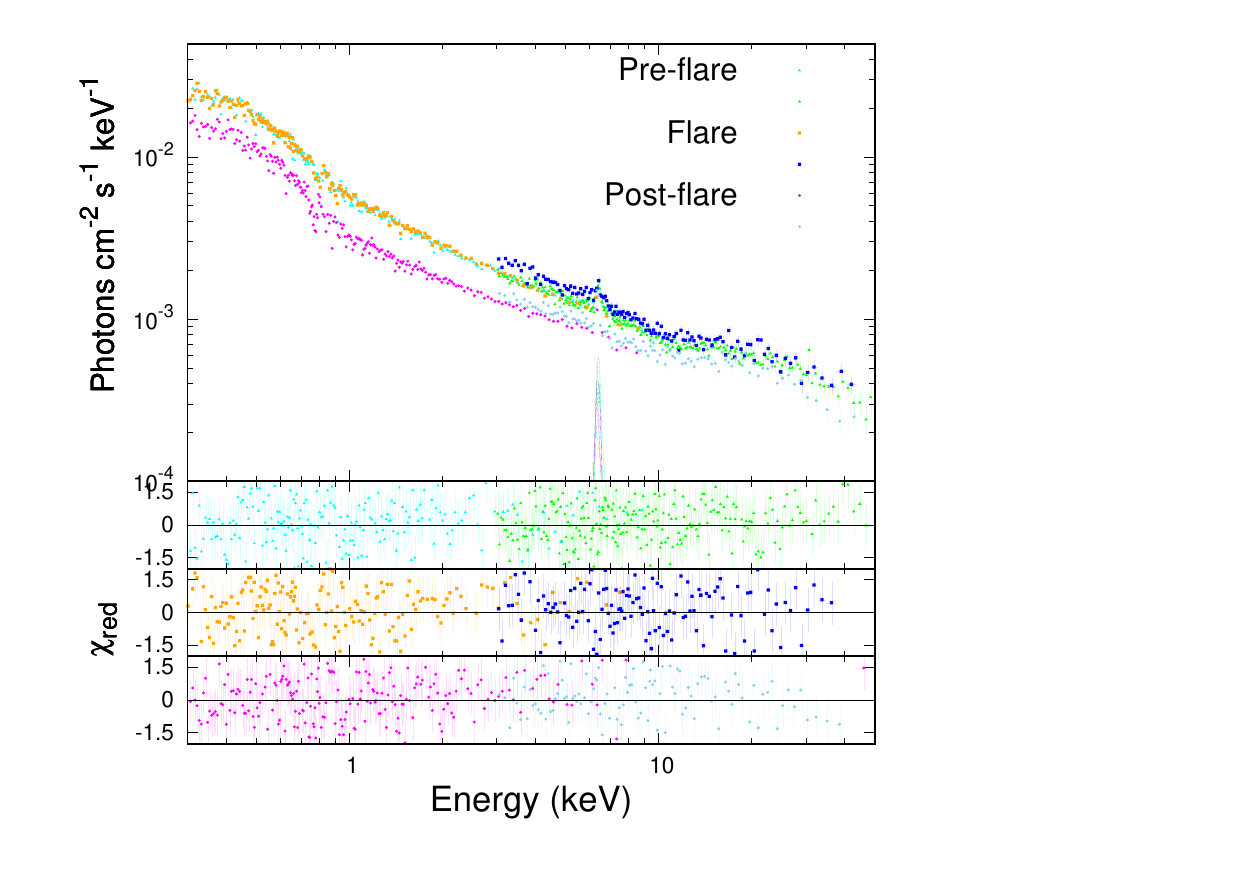}
         \caption{CompPS}
         \label{fig:compPS}
     \end{subfigure}
     \hspace{0.2in}
     \begin{subfigure}[b]{0.3\textwidth}
         \centering
         \includegraphics[width=1.7\textwidth]{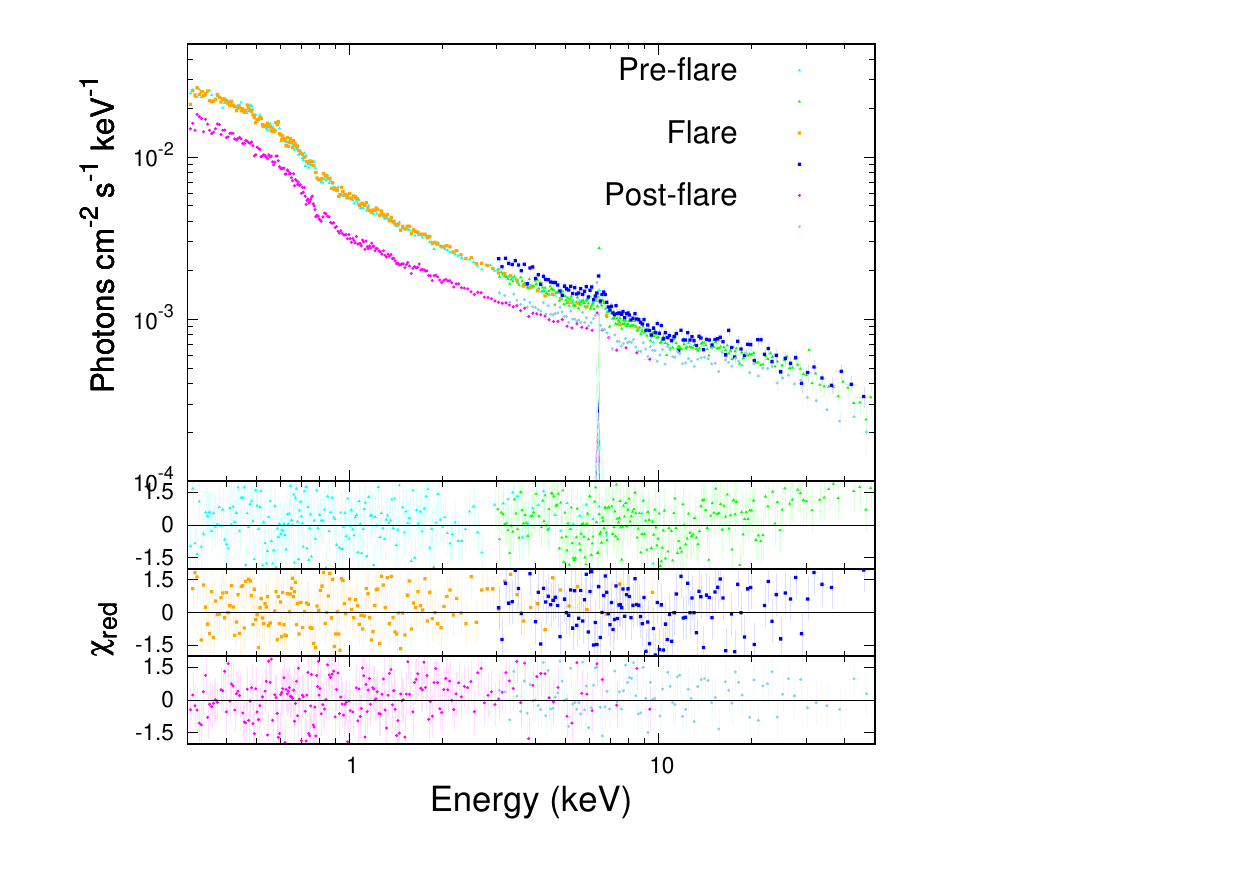}
         \caption{Relxilllpion}
         \label{fig:relxillpion}
     \end{subfigure}
        \caption{Top panel: The $0.3-50$~keV {\it XMM-Newton} + {\it NuSTAR} spectra fitted with the Model-1 (left panel), Model-2 (middle panel), and Model-3 (right panel) for the pre-flare, flare, post-flare phases. Bottom three panels: Corresponding residuals for the pre-flare, flare, post-flare phases, respectively. The cyan, orange and magenta points represent the {\it XMM-Newton} data for the pre-flare, flare, post-flare phases, respectively. The green, blue and turquoise points represent the {\it NuSTAR} data for the pre-flare, flare, post-flare phases, respectively.}
        \label{fig:spectra}
\end{figure*}

We considered a spherical geometry of the plasma in this model. The inclination was fixed at 30 degrees. The values of $\rm R_{in}$ and $\rm R_{out}$ were fixed at 6 and 2000 $\rm R_s$, respectively. The parameters obtained from fitting the data from three segments are tabulated in Table~\ref{tab:compPS}. Instead of the optical depth, this model incorporates Compton y-parameter ($\tau_y$) as one input which relates to the optical depth as $\tau=\tau_y/(4kT_e/m_e c^2)$, where $kT_e$ is the hot electron temperature in keV. Iron and heavy element abundances were tied for the time intervals. The absolute value of Fe K$\alpha$ line energy marginally varies during the three phases while the equivalent width (EW) is lowest in the flare phase. Inner disc temperature is found to be invariable (within error) with an average temperature of $\sim$18 eV. From this model, interestingly, we found that the coronal temperature increased to $\sim$228 keV post-flare while it was similar before and during the flare with a value of around $\sim$182 keV. Moreover, the reflection fraction was found to be decreasing during the flare while it substantially increased after the flare.
\begin{table}
\caption{The best-fitted parameters obtained after fitting 0.3-50 keV time-resolved spectra simultaneously using XMM-Newton and NuSTAR observations with phenomenological model {\tt compPS}.}
\begin{tabular}{lccc}
\hline
Parameters & Pre-flare & Flare & Post-flare \\
				           &segment &segment &segment\\
		\hline
		$N_{\rm H,1}$ ($10^{22}$ \cs) & $8.22_{-0.87}^{+0.25}$ & $7.55_{-1.47}^{+1.17}$ & $5.64_{-0.26}^{+0.36}$ \\
		\\
		$\log {\rm \xi_{\rm 1}}$ & $1.93_{-0.02}^{+0.09}$ & $1.89_{-0.04}^{+0.07}$ &$1.80_{-0.06}^{+0.02}$ \\
		\\
		Cov Frac1 & $0.35_{-0.02}^{+0.01}$ & $0.28_{-0.05}^{+0.07}$ & $0.45_{-0.02}^{+0.03}$ \\
		\\
		z2 & $-0.26_{-0.01}^{+0.01}$ & $-0.26_{-0.02}^{+0.02}$ & $-0.21_{-0.01}^{+0.01}$ \\
		\\
		$N_{\rm H,2}$ ($10^{22}$ \cs) & $0.47_{-0.08}^{+0.43}$ & $1.37_{-0.76}^{+2.02}$ & $1.24_{-0.54}^{+0.32}$ \\
		\\
		$\log {\rm \xi_{\rm 2}}$ & $2.15_{-0.06}^{+0.08}$ & $2.14_{-0.09}^{+0.13}$ & $2.04_{-0.05}^{+0.08}$ \\
		\\ 
		Cov Frac2 & $>0.63$ & $>0.47$ & $0.71_{-0.06}^{+0.16}$ \\
		\\
		z1 & $0.001_{-0.01}^{+0.01}$ & $0.01_{-0.01}^{+0.02}$ & $0.04_{-0.01}^{+0.01}$  \\
		\\
		Fe K$\alpha$ ~LE (keV) & $6.40_{-0.03}^{+0.02}$ & $6.37_{-0.05}^{+0.04}$ & $6.35_{-0.02}^{+0.02}$ \\
		\\
		$\sigma$ (eV) & $97_{-49}^{+55}$ & $115_{-58}^{+88}$ & $70_{-41}^{+38}$ \\
		\\
		EW (eV) & $125_{-17}^{+18}$ & $101_{-22}^{+25}$ & $118_{-18}^{+19}$ \\
		\\
		Norm ($10^{-05}$) & $2.38_{-0.33}^{+0.35}$ & $1.82_{-0.39}^{+0.44}$ & $1.56_{-0.23}^{+0.24}$\\
		\\ 
		$kT_{\rm e}$ (keV)& $182_{-23}^{+24}$ &  $184_{-26}^{+28}$ & $228_{-15}^{+52}$ \\
		\\
		$kT_{\rm bb}$ (eV)& $19_{-3}^{+3}$ &  $19_{-4}^{+5}$ & $16_{-3}^{+4}$ \\
		\\
		$\tau_y$ & $0.37_{-0.02}^{+0.02}$ &  $0.36_{-0.03}^{+0.03}$ & $0.31_{-0.01}^{+0.02}$ \\
		\\
		relrefl & $2.19_{-0.16}^{+0.16}$ & $ 2.10_{-0.23}^{+0.21}$ & $2.58_{-0.46}^{+0.29} $ \\
		\\
		Abund & $4.68_{-0.41}^{+0.47}$ & $\rm f$ & $\rm f$ \\
		\\
		Abund$_{\rm Fe}$ & $2.25_{-0.30}^{+0.29}$ & $\rm f$  & $\rm f$\\ \\
		\hline
\end{tabular}
\leftline{$\rm f$ represents the parameter fixed at the corresponding value at the pre-flare phase.}
\label{tab:compPS}
\end{table}

\begin{table}
\caption{The best-fitted parameters obtained after fitting 0.3-50 keV time-resolved spectra simultaneously using XMM-Newton and NuSTAR observations with {\tt bmc} model.}
\begin{tabular}{lccc}
\hline
Parameters & Pre-flare & Flare & Post-flare \\
				           &segment &segment &segment\\
		\hline
		$kT$ (eV)& $9.44_{-0.10}^{+0.10}$ &  $9.30_{-0.15}^{+0.15}$ & $9.28_{-0.12}^{+0.11}$ \\
		\\
		alpha & $1.10_{-0.01}^{+0.01}$ &  $1.15_{-0.01}^{+0.01}$ & $1.10_{-0.01}^{+0.01}$ \\
		\\
		logA & $0.05_{-0.01}^{+0.01}$ &  $0.20_{-0.02}^{+0.02}$ & $0.13_{-0.01}^{+0.01}$ 
            \\
		\\
		Norm$_{bmc}$ $ (10^{-04})$ & $4.30_{-0.02}^{+0.02}$ & $ 4.12_{-0.02}^{+0.02}$ & $2.61_{-0.01}^{+0.01} $ \\
		\\
		$kT_{hump}$ (keV) & $9.12_{-0.47}^{+0.53}$ & $9.68_{-0.81}^{+0.97}$ & $7.69_{-0.46}^{+0.51}$ \\
		\\
		Norm$_{hump}$ $ (10^{-04})$ & $1.97_{-0.14}^{+0.16}$ & $1.77_{-0.21}^{+0.25}$  & $1.46_{-0.12}^{+0.13}$\\ \\
		\hline
\end{tabular}
\leftline{$kT_{hump}$ is the temperature of the blackbody component used for the hump}
\leftline{above 10 keV.}
\label{tab:bmc}
\end{table}

We also tried the {\tt bmc} model \citep{1997ApJ...487..834T, 1998ApJ...493..863T, 1999ApJ...511..289L} for fitting the data. This model describes the spectrum generated not only by the thermal Comptonization but also by the Comptonization of soft photons by the matter going through the relativistic bulk motion. This is the self-consistent convolution model of power law and seed blackbody soft photons. The resultant spectrum is characterized by spectral index, $\alpha$ ($=\Gamma -1$), seed photon temperature ($kT$), and an illumination parameter $(A)$ which is the fractional illumination of bulk-motion flow by the thermal photons. The composite model in {\tt XSPEC} reads as
\begin{multline*}
	\tt constant\times tbabs \times zxipcf \times zxipcf \times (bmc+zgauss+zbbody)
\end{multline*}

Here, we used {\tt zbbody} to fit the high energy hump above 10 keV. The fitted model parameters are listed in Table~\ref{tab:bmc}. 

Using the correlation of the photon index and {\tt bmc} normalization, the mass of the black hole can be estimated. This concept has been discussed in detail and implemented for NGC~4051 in \cite{2018A&A...613A..48S}. The expression used by \cite{2018A&A...613A..48S} to calculate the mass of the black hole is given by

$$M_{\rm t}=M_{\rm r} \frac{N_{\rm t}}{N_{\rm r}} \Big( \frac{d_{\rm t}}{d_{\rm r}}\Big)^2 f_{\rm G}$$

where $M$, $N$, \& $d$ are the black hole mass, normalization and distance of the source, respectively. In this equation, the subscript $t$ represents the target source, whereas $r$ represents the reference source. Here, $f_{\rm G}=cos(\theta_{\rm r})/cos(\theta_{\rm t})$ is the geometrical factor which takes into consideration the difference in the viewing angles of the inner disc from where the seed photons are being emitted for both the target and reference sources. $\theta$ is the inclination angle of the inner disc which has been taken equal to the inclination angle of the source. Here, we consider NGC~4051 as the target source and GRO~J1655-40 as the reference source. Referring to the \cite{2018A&A...613A..48S} estimations of GRO~J1655-40, $N_{\rm r}=7\times10^{-2}$, $\theta_{\rm r}=70$ degrees and $d_{\rm r}=3.2$ kpc and using our estimate of normalization for NGC~4051 as $N_{\rm t}\approx4\times10^{-4}$ and $\theta_{\rm t}$ = 30 degrees, we calculate the black hole mass in NGC~4051 as $\ge 1.32\times 10^5 M_{\odot}$ for a source distance of $d_{\rm t}=9800$ kpc \citep{2018A&A...613A..48S}. Our estimated value of black hole mass in NGC~4051 is lower than that estimated by \cite{2018A&A...613A..48S} ($6.1\times 10^5 M_{\odot}$). The discrepancy in the estimated values of the black hole mass in NGC~4051 is due to the difference in the values of $f_{\rm G}$ and $N_{\rm t}$. \cite{2018A&A...613A..48S} used $f_{\rm G}$ and $N_{\rm t}$ as 1 and $\approx8.1\times10^{-3}$, whereas from our analysis, the value of $f_{\rm G}$ is estimated to be 0.39.

{\bf Model-3:} Reflection from the photoionized accretion disc is considered to be one of the explanations of the origin of the soft X-ray excess in the AGNs. Emissions from the compact, hot and relativistic plasma irradiate the accretion disc. The X-ray illumination is stronger in the inner regions of the accretion disc due to the strong gravity of the central black hole. We used the {\tt relxilllpion} variant of the {\tt relxill} model to describe the soft X-ray excess. The {\tt relxill} model computes the continuum emission and its reflection from the accretion disc that includes the reprocessing of the continuum photons in the disc. The blurring of the soft X-ray emission lines due to the relativistic motion of the inner disc gives a smooth curvature, which appears as soft X-ray excess. This model is parameterized by the inclination angle, inner radius, reflection fraction and ionization parameter ($\rm log\xi$) with constant ionization throughout the disc. The {\tt relxilllpion} model allows the ionization of the disc to vary with radius, and this variation is approximated by power-law. 
The composite model in {\tt XSPEC} reads as
\begin{multline*}
	\tt constant\times tbabs \times zxipcf \times zxipcf \times (relxilllpion)
\end{multline*}

While fitting the data with this model, we tied the iron abundance across the three intervals. The Eddington ratio for this source is calculated to be $\sim$0.01 for the black hole mass of $1.73\times10^6$ M$_{\odot}$ and bolometric luminosity $\sim 2.3\times10^{42}$\ergsec. The cutoff energy is kept fixed at 370 keV as earlier. We required two absorbers to fit the residuals in the soft X-ray range ($<2$ keV) of the spectra. Initially, we varied the value of inner disc radius ($\rm R_{in}$) and ionization parameter ($\rm log\xi$) and found fairly unchanged parameter values. Therefore, we tied these parameters for all three segments. The height of the corona from the disc increased to $12.18_{-0.92}^{+0.98}$ R$_g$, accompanied by a reduction in the reflection fraction during the flare. After the flare, the coronal height decreased with the highest reflection fraction among the three phases. 

\begin{table}
\caption{The best fitted parameters obtained after fitting $0.3-50$ keV time-resolved spectra simultaneously using {\it XMM-Newton} and {\it  NuSTAR} observations with physical model {\tt relxilllpion}.}
\label{tab:relxillpion}
\begin{tabular}{lccc}
\hline
Parameters & Pre-flare & Flare & Post-flare \\
           & segment   &segment &segment\\
\hline
$N_{\rm H,1}$ ($10^{22}$ \cs) & $21.04_{-2.39}^{+4.79}$ & $46.46_{-4.09}^{+4.89}$ & $47.59_{-9.00}^{+6.93}$ \\
\\
$\log {\rm \xi_{\rm 1}}$ & $1.92_{-0.07}^{+0.08}$ & $2.00_{-0.03}^{+0.03}$ & $1.77_{-0.30}^{+0.23}$ \\
\\ 
Cov Frac1 & $0.27_{-0.02}^{+0.02}$ & $0.30_{-0.01}^{+0.01}$ & $0.30_{-0.01}^{+0.01}$ \\
\\
z1 & $-0.33_{-0.02}^{+0.01}$ & $-0.31_{-0.02}^{+0.03}$ & $-0.36_{-0.03}^{+0.02}$  \\
\\

$N_{\rm H,2}$ ($10^{22}$ \cs) & $3.98_{-0.59}^{+0.59}$ & $5.44_{-0.47}^{+0.39}$ & $1.07_{-0.30}^{+0.11}$ \\
\\
$\log {\rm \xi_{\rm 2}}$ & $1.91_{-0.06}^{+0.09}$ & $1.98_{-0.11}^{+0.08}$ &$1.90_{-0.07}^{+0.04}$ \\
\\
Cov Frac1 & $0.28_{-0.02}^{+0.03}$ & $0.35_{-0.03}^{+0.02}$ & $0.45_{-0.01}^{+0.01}$ \\
\\
z2 & $-0.01_{-0.01}^{+0.02}$ & $-0.01_{-0.02}^{+0.01}$ & $0.001_{-0.008}^{+0.007}$ \\
\\

h & $9.61_{-1.68}^{+1.38}$ & $12.18_{-0.92}^{+0.98}$ & $7.36_{-0.98}^{+1.51}$ \\
\\
a & $>0.85$ & $\rm f$ & $\rm f$ \\
\\
$R_{\rm in}$ & $<2.52$ & $\rm f$ & $\rm f$ \\
\\
$\log {\rm \xi_{\rm 1}}$ & $3.10_{-0.06}^{+0.05}$ & $\rm f$ & $\rm f$ \\
\\
$\Gamma$ & $2.16_{-0.01}^{+0.01}$ & $2.18_{-0.01}^{+0.01}$ & $2.15_{-0.01}^{+0.01}$ \\
\\
Abund$_{\rm Fe}$ & $1.02_{-0.04}^{+0.16}$ & $\rm f$ & $\rm f$ \\
\\
Refl. frac & $2.00_{-0.06}^{+0.06}$ & $1.42_{-0.05}^{+0.05}$ & $2.12_{-0.06}^{+0.07}$ \\
\\
Ionization index & $1.07_{-0.05}^{+0.08}$ & $1.15_{-0.09}^{+0.09}$ & $1.03_{-0.05}^{+0.05}$\\
\\
Norm$_{\rm PL}$ (10$^{-4}$) & $1.79_{-0.72}^{+0.71}$ & $1.91_{-0.05}^{+0.06}$ & $1.73_{-0.18}^{+0.19}$ \\
\\

\hline
\end{tabular}
	\leftline{$\rm f$ represents the parameter fixed at the corresponding value at the pre-flare phase.}
\leftline{}
\leftline{} 
\end{table}

\begin{table}
	\caption{Fluxes and Luminosities}
	\label{tab:lumin}
	\begin{tabular}{lccc}
		\hline
		Parameters & Pre-flare & Flare & Post-flare \\
		           &segment &segment &segment\\
  \hline

	    $F_{\rm 0.3-2~keV}^{\rm BB} (10^{-12})\star$ & $7.65\pm0.08$ & $5.88\pm0.09$  & $4.54\pm0.05$  \\ \\
	    $F_{\rm 2-10~keV}^{\rm PL} (10^{-11})\star $ & $1.77\pm0.01$ & $1.82\pm0.01$ & $1.26\pm0.06$   \\ \\
	   $L_{\rm 0.1-200}$ ($10^{41})\bullet$ & $8.55\pm0.03$ & $9.20\pm0.03$ & $6.04\pm0.02$ \\ \\
	    $L_{\rm ion}$ ($10^{41}) \bullet$ & $5.85\pm0.02$ & $6.18\pm0.02$ & $4.04\pm0.01$ \\ \\
	     $L_{\rm 2-10~keV}$ ($10^{41}) \bullet$ & $2.06\pm0.01$ & $2.39\pm0.01$ & $1.5\pm0.01$\\ \\
	     $\lambda_{\rm Edd}$ & $0.014\pm0.001$ & $0.016\pm0.001$ & $0.011\pm0.001$ \\ \\
	     $l$&$ 9.12_{-0.03}^{+0.02}$& $9.82_{-0.05}^{+0.02}$ & $6.44_{-0.02}^{+0.03}$ \\ \\
	     $\theta$&$ 0.36_{-0.04}^{+0.05}$& $ 0.36_{-0.06}^{+0.06}$ & $ 0.45_{-0.03}^{+0.10}$ \\
	    \hline
	  
	\end{tabular}
	\leftline{$\star$ in units of \funit, ~~~~~~~$\bullet$ in units of \ergsec} 
\end{table}

\begin{figure*} 
	\centering
	\includegraphics[scale=0.23, angle=-90 ,trim={0 0 0 0}, clip]{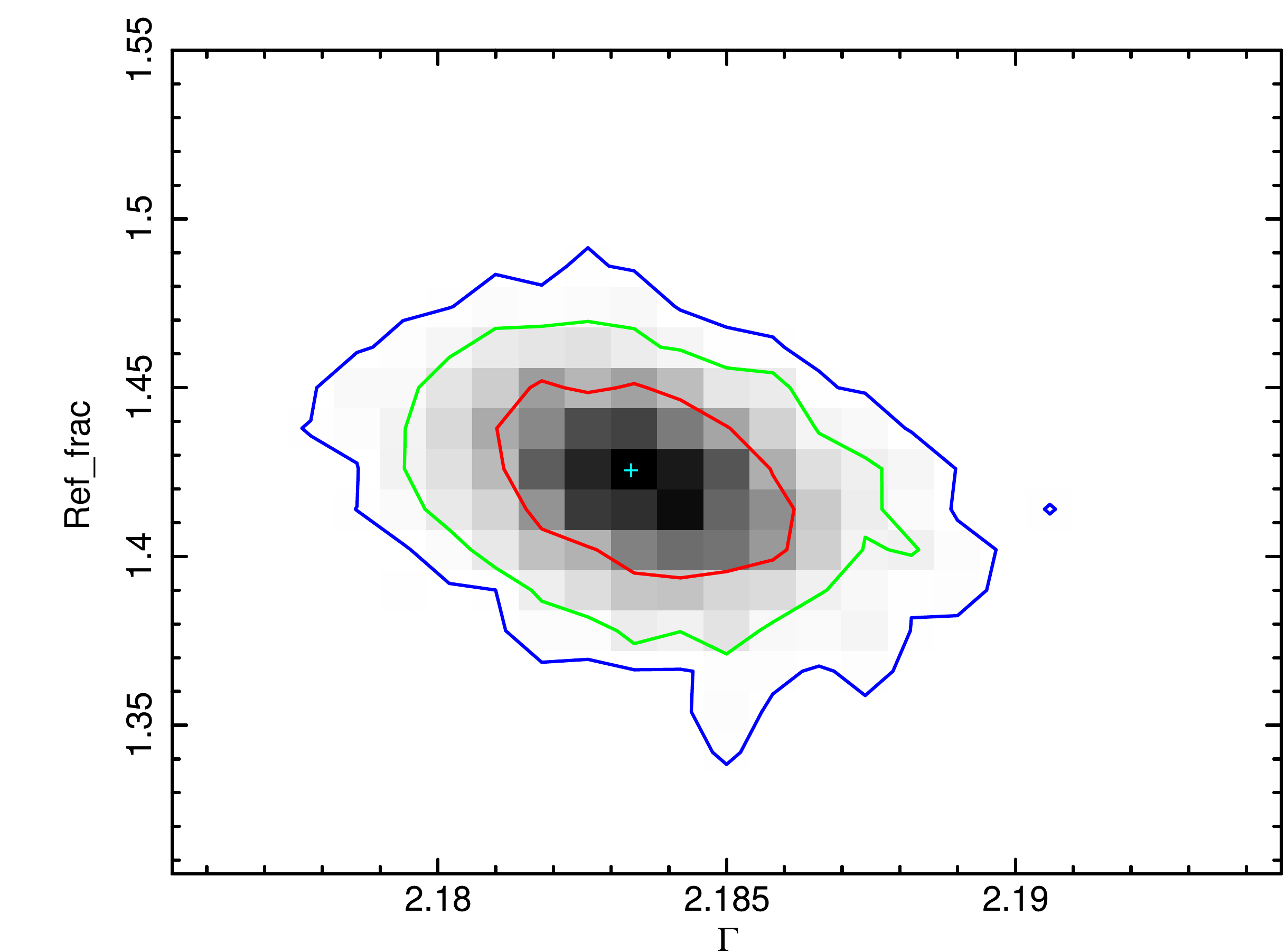} 
	\includegraphics[scale=0.23, angle=-90 ,trim={0 0 0 0}, clip]{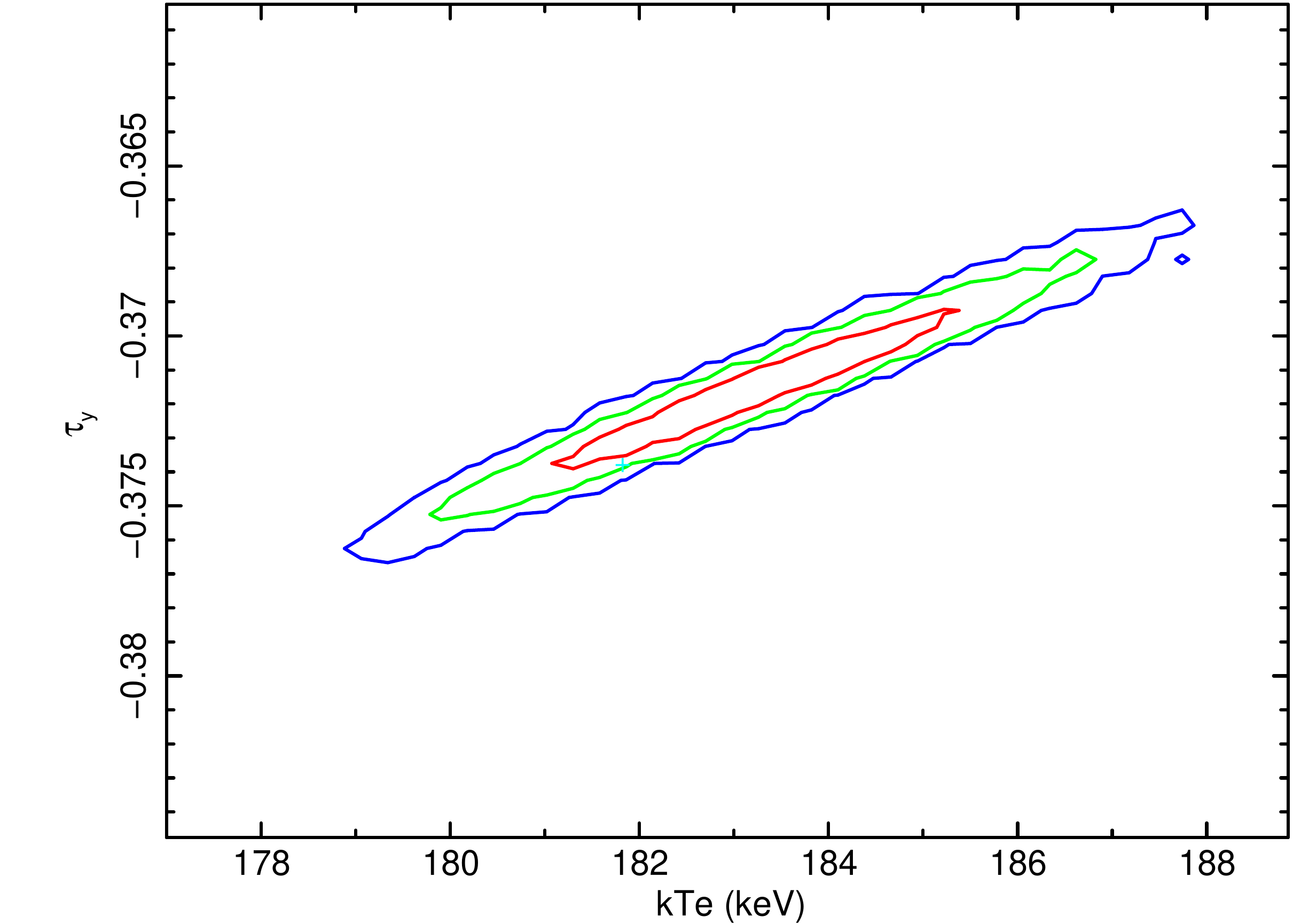}
	\includegraphics[scale=0.23, angle=-90 ,trim={0 0 0 0}, clip]{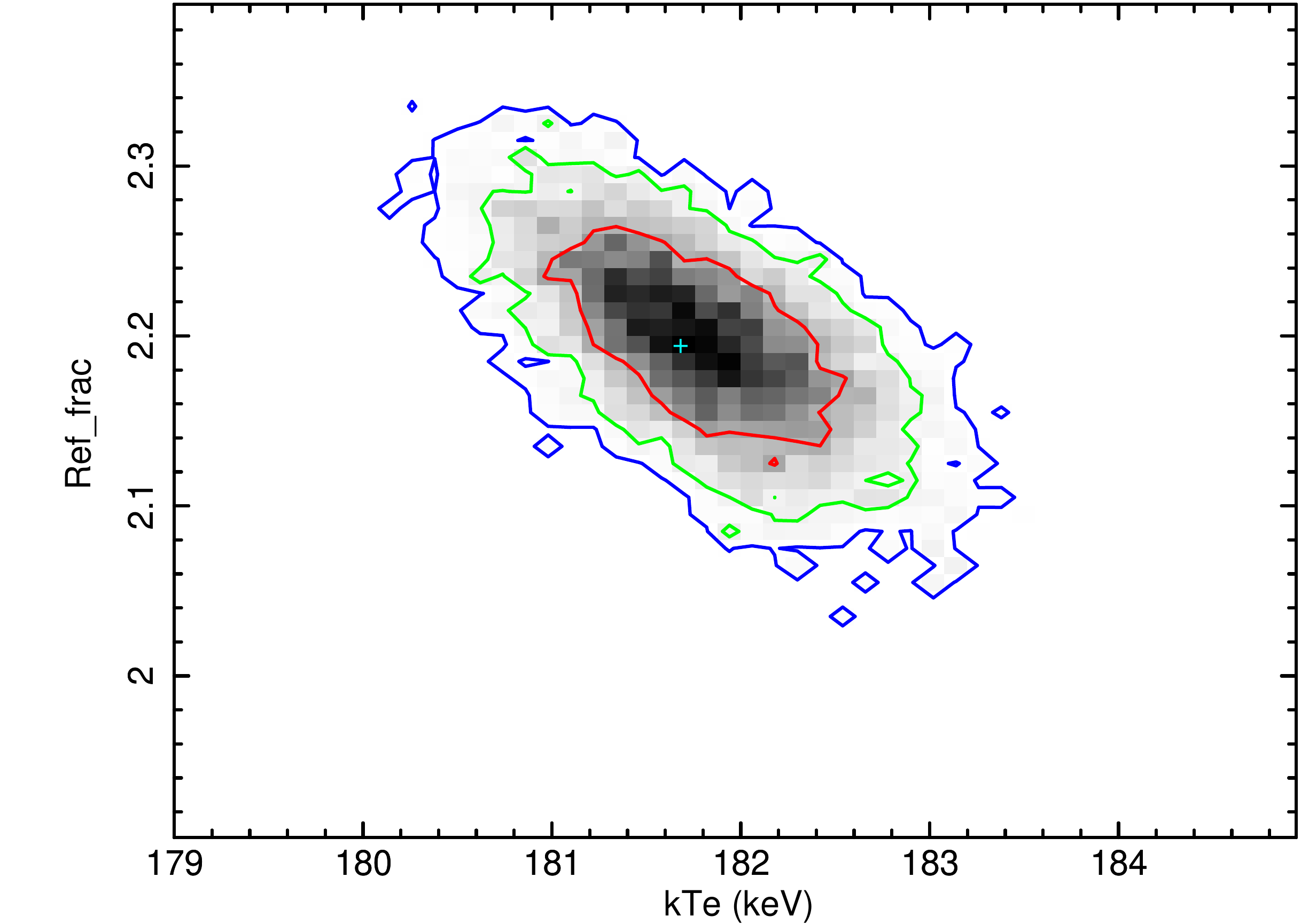} 
	\caption{Contour plots resulted from the MCMC analysis of the {\tt relxilllpion}, {\tt compPS} models for the reflection fraction vs $\Gamma$ (left panel), Compton y-parameter ($\tau_y$) vs $\rm kT_e$ (middle panel) and reflection fraction vs $\rm kT_e$ (right panel).}
	\label{fig:gamma_ref} 
\end{figure*}

\section{Discussion}
\label{sec:Discussion}

In this work, we explored the changes in the spectral and timing properties of NGC~4051 during the pre-flare, flare and post-flare phases using the data obtained from the \xmm~and \nustar~ observations in November 2018. We explored the accretion properties, accretion mechanism and the corona structural changes of the source in detail using various phenomenological and physical models. The results obtained from our work are described here.

\subsection{Black Hole spin}
The spin of the black hole is estimated self-consistently from the X-ray reflection models. In the present work, the spin parameter was linked while fitting the spectra of three segments simultaneously. From the relativistic reflection model {\tt relxilllpion}, we found that the black hole is rapidly spinning with the spin parameter $a>0.85$ while keeping the values of the inner radius and spin parameter tied for all three time intervals.  

\subsection{Location of the absorbers}
While fitting the spectra from the simultaneous observations of NGC~4051 with \xmm~ and \nustar, we required two absorbers to get rid of the residuals below 2 keV in all the spectral models used in this work. We took into account absorption with partially ionized and partially covered absorber models {\tt zxipcf}. The properties of these absorbers are listed in Tables~\ref{tab:zcutof}, \ref{tab:compPS}, \& \ref{tab:relxillpion}. The ionization parameter of an absorber is defined as $\xi= L_{\rm ion}/nr^2$ \citep{1969ApJ...156..943T}, where $L_{\rm ion}$, $n$ and $r$ are the unabsorbed ionizing luminosity of the emitting source, number density and radial location of the absorber from the central source, respectively. The absorption from the excited states or absorption variability can be used to determine the limits of $r$ using the definition of $\xi$. It is possible to estimate the upper limit of the radial distance based on the assumption that the thickness of the absorber can not exceed its location from the SMBH \citep{2005A&A...431..111B, 2012ApJ...753...75C}; 
$$r\le r_{\rm max}=\frac{L_{\rm ion}}{N_{\rm H}\xi}$$ 
where $N_{\rm H}$ is the column density of the absorber. 

A lower limit on the location of the outflow can be calculated as

$$r\ge r_{\rm min}= \frac{2GM_{\rm BH}}{\nu^2}$$
where $G$, $M_{\rm BH}$ and $\nu$ are the gravitational constant, the black hole mass and the outflow velocity, respectively.

Another way of calculating the lower limit is by considering the light travel time $D=\Delta t\times c$ during the observation if the absorber does not seem to be variable. We used the values from the {\tt zcutoffpl} model to estimate the radial location of both absorbers. The ionizing luminosity of this source has been calculated for three intervals by considering the power law component in the 0.3-50 keV range (see Table~\ref{tab:lumin}). The locations of the absorber~1 and absorber~2 are estimated and listed in Table~\ref{tab:zcutof} separately for three time intervals. These values of locations and redshift are consistent with the  WAs and UFOs found in the studies of type~1 Seyfert galaxies \citep{2012MNRAS.422L...1T, 10.1093/mnras/sts692}.

\begin{figure*} 
	\centering
	\includegraphics[scale=0.4, angle=0 ,trim={0 0 0 0}, clip]{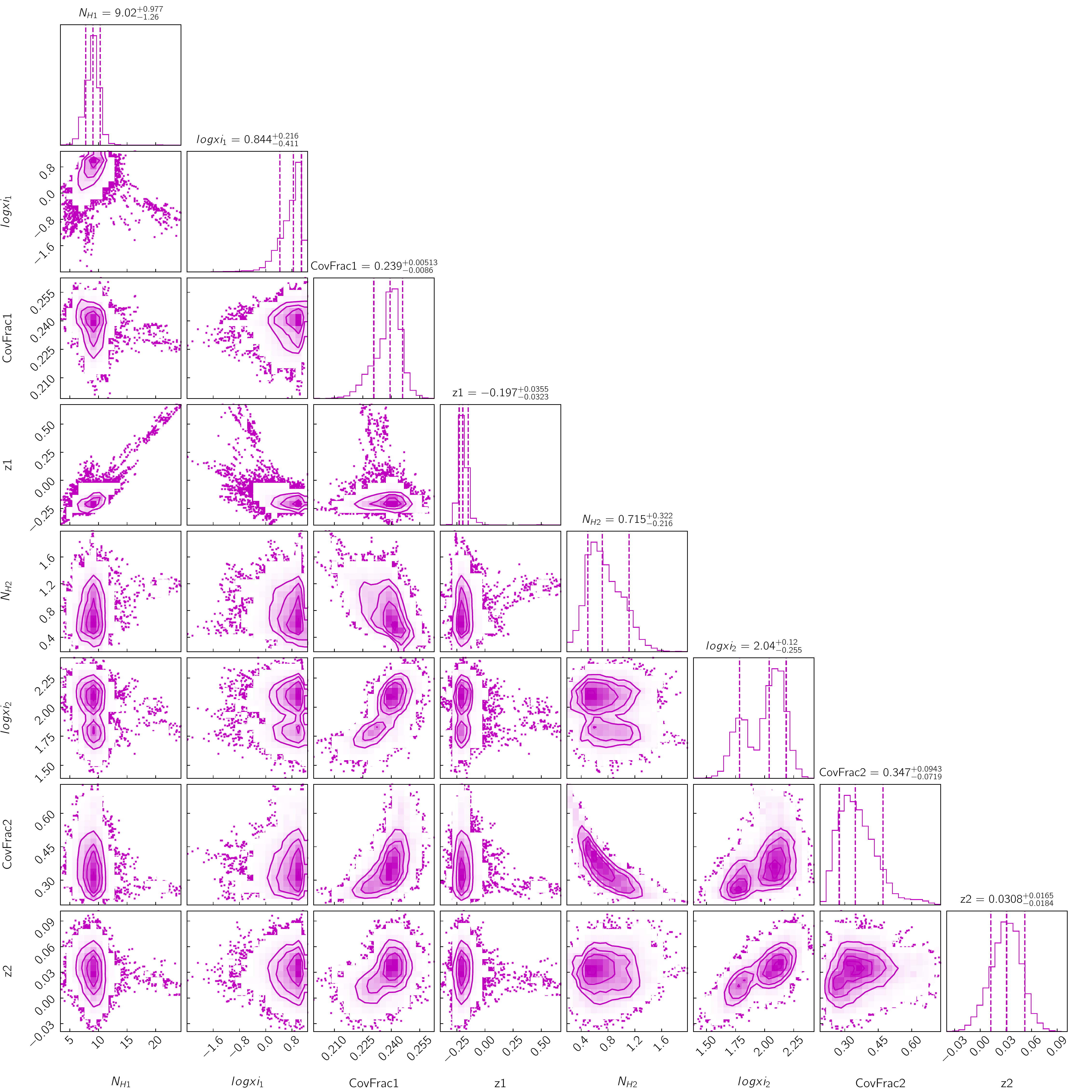}
	\caption{The 1D and 2D posterior distribution curves resulted from the MCMC analysis of the ionization parameter ($log\xi$), covering factor, and redshift ($z$) of the two absorbers. The vertical lines in the 1D distribution show 16\%, 50\%, and 90\% quantiles. CORNER.PY \citep{corner} was used to plot the distributions. }
	\label{fig:mcmc} 
\end{figure*}

We have also calculated the marginal posterior distributions for the absorber parameters to check the spectral fitting degeneracy between them. To search through the parameter space, we used the Markov Chain Monte Carlo (MCMC) sampling procedure. We used \citet{2010CAMCS...5...65G} algorithm implemented in {\tt XSPEC} to determine the chain. We considered 20 walkers with a total chain length of 200000 and burnt the initial 20000 steps. The one-dimensional (1D) and two-dimensional (2D) distributions are shown in Figure~\ref{fig:mcmc}.

\subsection{Properties of corona}

The X-ray emitting corona is characterized by the optical depth ($\tau$), hot electron plasma temperature ($\rm kT_e$) and the X-ray continuum photon index ($\Gamma$). As the power-law with high energy cut-off continuum model gives information only on the photon-index, we utilized the {\tt compPS} model, which considers $\rm kT_e$, Compton $y$ parameter and inner disc temperature ($kT_{\rm bb}$) as free parameters. 
We find that the temperature of the corona increased after the flare subsided and the source went into the low flux state. During the flare, when flux was high, the slope of the spectrum became steep. This softening of the X-ray continuum with high flux is consistent with the "softer when brighter" nature observed in many AGNs \citep{2016MNRAS.459.3963C}. The less energetic seed photons from the accretion disc get Comptonized in the corona, producing X-ray emission. In the present work, it is found that the coronal temperature increased after the flare. An increase in the temperature of the corona after the flare can be interpreted as due to the injection of fewer seed photons into the corona resulting in the reduction of the Compton cooling rate. Considering the magnitude of the observed flare and the duration of the observation, it is highly unlikely that there could be any significant variation in the accretion rate (see $\rm \lambda_{Edd}$ values from Table~\ref{tab:lumin}). Moreover, the flux of the soft X-ray excess ($0.3-2$ keV range) does not show any clear correlation with the photon index (see Table~\ref{tab:lumin}). This indicates that the change in the spectral state during the flaring phase is inherent to the corona instead of the variations in the seed photons. In an AGN, the corona is likely to be powered by the small-scale magnetic flux tubes associated with the orbiting plasma in the accretion disc \citep{2019MNRAS.487.4114Y}. Any change in the magnetic field can cause a change in the location and geometry of the corona, which in turn can result in coronal temperature variation. Recently, \cite{10.1093/mnras/stac416} studied a  similar kind of X-ray flaring event in NLS1 galaxy I Zw 1. They found that the temperature was higher in the pre-flare phase and decreased during the flare, probably due to the expansion of the corona accompanied by softening of the spectrum. After the flare, the temperature again started to increase with the hardening of the continuum spectrum. In NGC~4051, we found that the coronal temperature increased after the flare and the spectrum became harder. In this case, the corona perhaps started inflating in the pre-flare phase and reached a maximum during the flare. After the flare, it again contracted, followed by an increase in temperature.
    
\begin{figure} 
	\centering
	\includegraphics[scale=0.28, angle=0 ,trim={0.1in 0 0 0}, clip]{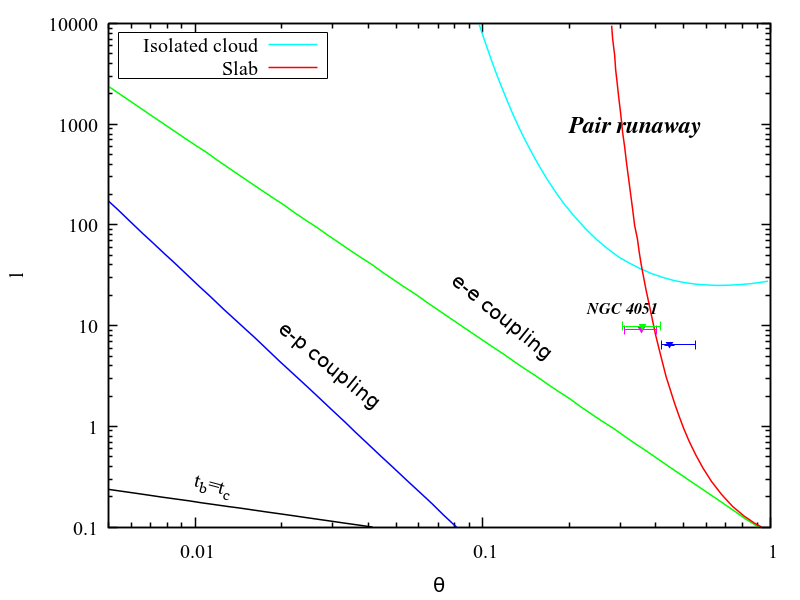}
	\caption{Theoretical compactness-temperature diagram is shown. The green line represents the region below which the electron-electron coupling time scale is shorter than the Compton cooling time while the region below the blue line, the electron-proton coupling time is shorter. Bremsstrahlung is dominant in the region below the black line. The region beyond the cyan and red curves represents where pair run-away process occur for isolated cloud \citep{1984MNRAS.209..175S} and slab geometry \citep{1995ApJ...449L..13S}. The green, magenta and blue points are drawn for pre-flare, flare and post-flare phase positions of NGC~4051, respectively.}
	\label{fig:compactness} 
\end{figure}

In the present work, the coronal temperature was found to be high (average value of $\sim$198 keV). This is likely the reason we could not constrain the cut-off energy for this source. From the best-fitted values of the coronal temperature ($kT_e$) and Compton y-parameter ($y$) (although these two parameters are degenerate as shown in the middle contour plot of Figure~\ref{fig:gamma_ref}), the value of optical depth ($\tau=\frac{\tau_y}{4T_e/m_e c^2}$) is calculated as $0.26\pm0.04$, $0.25\pm0.04$, $0.17\pm0.03$ in the pre-flare, flare and post-flare phases, respectively. These values of $\tau$ are consistent with the mean value obtained for a sample of 838 Swift/BAT AGNs \citep{2018MNRAS.480.1819R}. The coronal temperature carries important information on the state of the X-ray emitting plasma. The position of the AGN in the compactness-temperature ($\theta-l$ plane) space provides an insight to the dominant process in the corona \citep{2015MNRAS.451.4375F}. In the corona of an AGN, the most significant processes are inverse Compton scattering, bremsstrahlung and pair production. Among these, the dominant process is the one for which the cooling time is the shortest. The Comptonization process dominates when $l>3\alpha_f \theta^{-1/2}$, where $\alpha_{\rm f}$ and $\theta$ are the fine structure constant and coronal temperature normalized to the electron rest energy, respectively. The  electron-proton coupling dominates when $3\alpha_{\rm f} \theta < l< 0.04 \theta^{-3/2}$ while for $0.04\theta^{-3/2} < l < 80\theta^{-3/2}$, the electron-electron coupling becomes prominent. There is a regime in the $\theta-l$ plane where the pair production process becomes the runaway process that depends on the corona shape and on the radiation mechanism. \citep{1995ApJ...449L..13S} computed the position of these pair runaway lines for a slab corona. \citep{1984MNRAS.209..175S} calculated this regime for isolated clouds and gave the condition for its occurrence when $l\sim 10 \theta^{5/2} e^{1/\theta}$. The compactness parameter is defined as,
    $$l=\frac{L}{R} \frac{\sigma_{\rm T}}{m_{\rm e} c^3}$$ 
where $L$, $R$, $\sigma_{\rm T}$, $m_{\rm e}$ and $c$ are the luminosity, radius of the corona, the Thomson cross-section, mass of the electron and speed of light, respectively. We estimated this parameter for NGC~4051 considering power-law continuum luminosity extrapolated to the $0.1-200$ keV band in three intervals and radius as 10~$R_g$ (for standard sources, \citep{2015MNRAS.451.4375F}). The $l$ and $\theta$ values are given in Table~\ref{tab:lumin} and plotted in Figure~\ref{fig:compactness}. NGC~4051 is found to be located at the edge of the pair runaway region corresponding to the slab corona. This suggests that the process is mostly dominated by pair production and annihilation. In a compact corona ($l>1$) when the photons are highly energetic, the photon-photon collision generates an electron-positron pair. If the temperature reaches $\sim$1 MeV, the pair production process becomes significant, thereby soaking energy and limiting the further rise in temperature.

\subsection{Soft X-ray excess and reflection} The soft X-ray excess above the continuum is the most common feature seen in the AGNs. This component is prominent in the NLS1s. However, the origin of the soft X-ray excess is still part of the discourse. There are various models in the literature that take account of the origin of this component. The Comptonization of the photons in an optically thick and warm corona in the inner part of the disc is one explanation \citep{10.1111/j.1365-2966.2011.19779.x} for the origin of the soft excess. Another scenario is the one in which the photons from the hot corona ($\sim$100 keV) get reflected from the relativistically moving inner accretion disc and give rise to the soft excess \citep{10.1046/j.1365-8711.2002.05419.x, 2005MNRAS.358..211R, 2016MNRAS.457..875P}. In the ionized absorption model, the photons are absorbed by the high-velocity winds originating from the accretion disc and re-emitted \citep{2004MNRAS.349L...7G}. 

In our analysis, we initially used the {\tt zbbody} model to fit the soft X-ray excess. We found the blackbody temperature ($\rm kT_{\rm BB}$) to be  $\sim$ 100 eV which was almost constant throughout the observation. This non-variation or minimal variation of the soft excess temperature is expected, as suggested by earlier studies \citep{2004MNRAS.349L...7G, 2017ApJS..233...17R}. We calculated the unabsorbed flux of the blackbody component for the soft excess and found it to be unaffected during the flare and decreasing with time. 

\begin{figure} 
	\centering
	\includegraphics[scale=0.3, angle=0 ,trim={-0.4in 0 0 1.0}, clip]{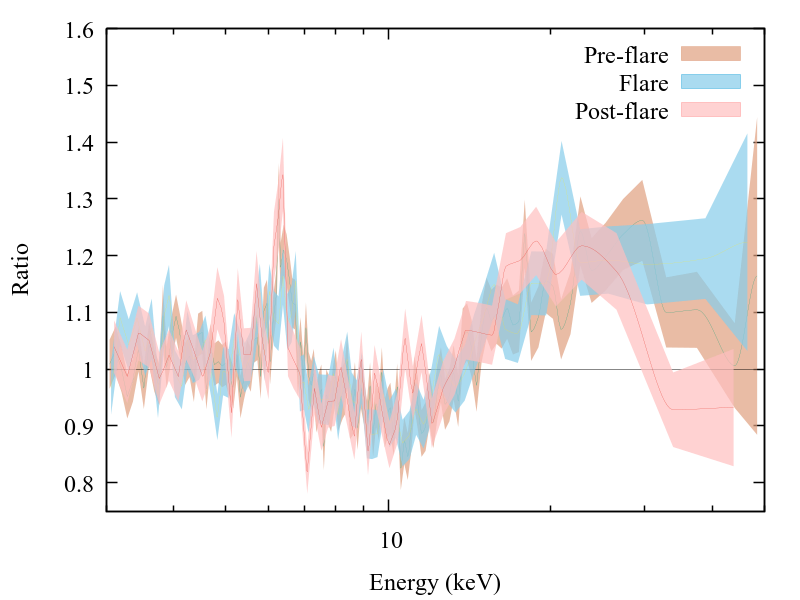}
	\caption{The ratio of the 3-50 keV spectrum from \nustar~ observation and best fitted with power-law model is shown. The varying strength of the reflection component above 10 keV is evident for all three phases of the observation.}
	\label{fig:reflection} 
\end{figure}

\begin{figure} 
	\centering
	\includegraphics[scale=0.5, angle=0 ,trim={0.4in 0 0 1.0}, clip]{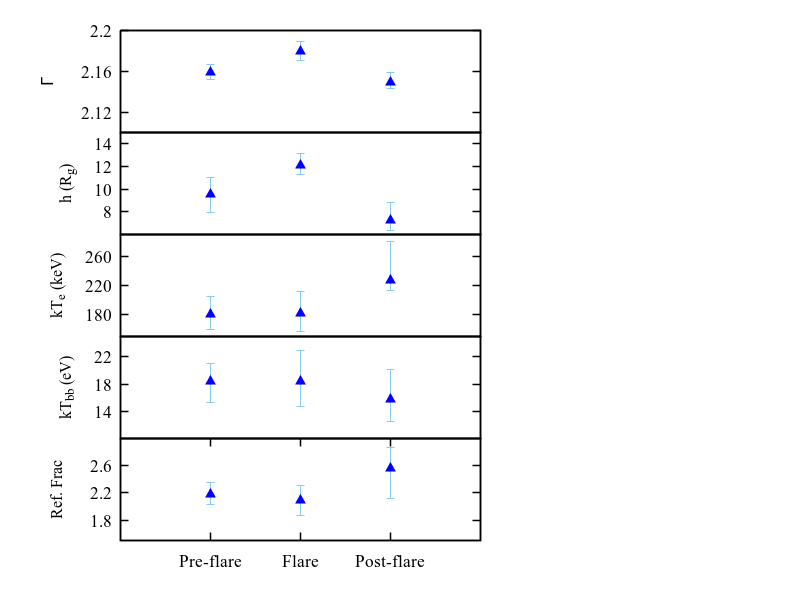}
	\caption{Variation of the best fit parameters are shown for three phases. The photon-index ($\Gamma$), coronal height (h) and reflection fraction (R) are obtained from {\tt relxilllpion} model while coronal temperature ($kT_{\rm e}$) and inner disc temperature (kT$_{\rm bb}$) has been taken from the {\tt compPS} model.}
	\label{fig:parameter} 
\end{figure}

From the {\tt relxilllpion} model, we found that during the flare, the reflection fraction decreased to $1.42$ from $2.00$ in the pre-flare phase and again increased to $2.09$ in the post-flare phase (see bottom panel of Figure~\ref{fig:parameter}). Similar trend was also observed from the {\tt compPS} model. The change in the reflection component can be seen in Figure~\ref{fig:reflection}, where the ratio between the 3-50 keV spectrum from the \nustar~observation and the best-fitted power-law is plotted. The reflection fraction is the ratio of reflected emission from the disc and continuum emission from the corona observed directly. From our analysis, we found this factor to be greater than unity, which means the source is reflection dominated. The gravitational bending causes the X-rays to be concentrated towards the accretion disc resulting in a high reflection fraction \citep{2004MNRAS.349.1435M}. As we mentioned above, NGC~4051 hosts a highly spinning black hole. Therefore, the effect of light bending causing a high reflection fraction is expected. In the simple lamppost geometry scenario, the compact corona is located above the central black hole. The drop of reflection fraction during the flare can be explained in terms of the coronal height from the accretion disc. We found that the coronal height increased, i.e., the corona moved away from the accretion disc during the flare when the reflection fraction was lowest (second from the top and bottom panels of Figure~\ref{fig:parameter}). After the flare, coronal height decreased, and the reflection fraction increased again. Another possibility of the drop in the reflection during the flare and enhancement after the flare could be due to the change in the inner accretion disc radius \citep{2021A&A...654A..14D}. As mentioned before, we did not find any significant change in the inner radius during three phases. Over-ionization could be another reason for the drop of the reflection during the flare. However, the ionization parameter was invariable during the observation and was fixed for the final fitting. Therefore, the motion of the corona from the accretion disc seems to be the most plausible explanation for the variable reflection fraction. 

\begin{figure}
     \begin{subfigure}[b]{0.3\textwidth}
          \caption{Pre-flare segment}
         \includegraphics[scale=0.7, angle=0 ,trim={0 1in 0 0}, clip]{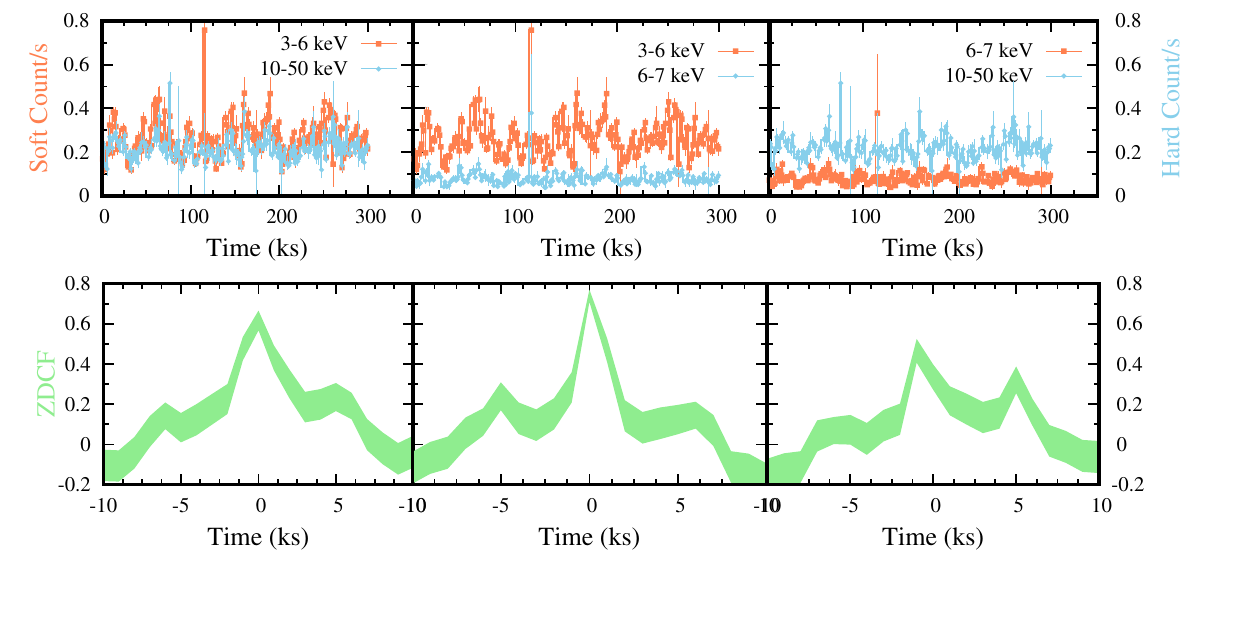}
         \label{fig:pre-flare}
     \end{subfigure}
     \hfill
          \begin{subfigure}[b]{0.3\textwidth}
         \caption{Flare segment}
         \includegraphics[scale=0.7, angle=0 ,trim={0 1in 0 0}, clip]{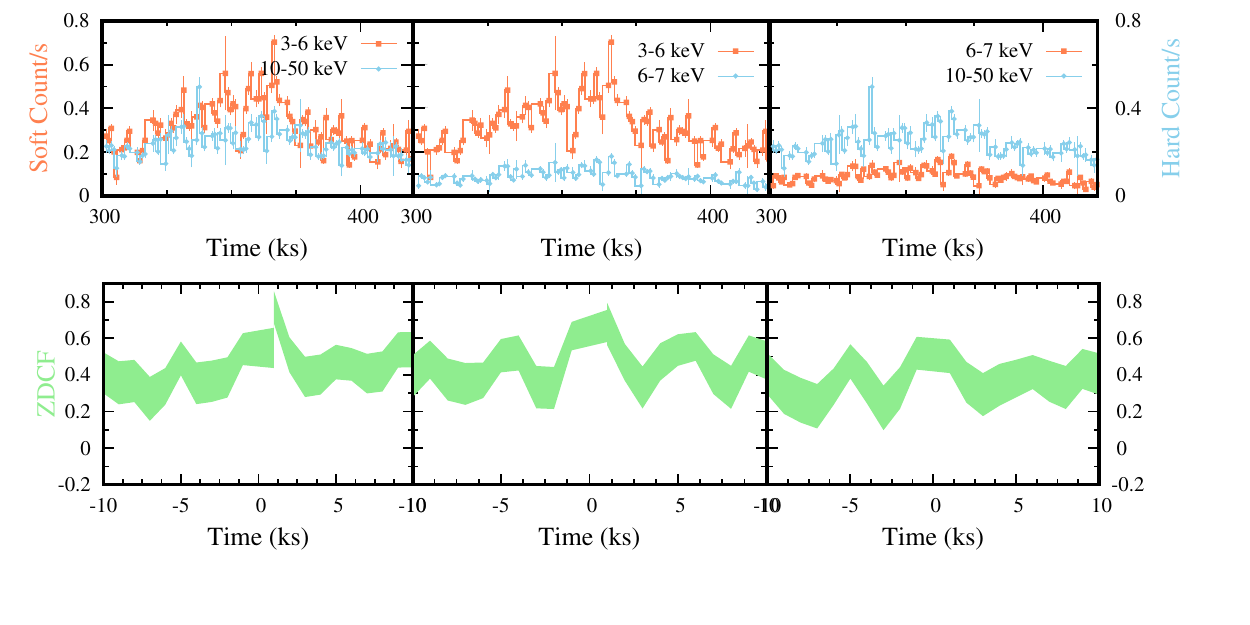}
         \label{fig:flare}
     \end{subfigure}
     \hfill
     \begin{subfigure}[b]{0.3\textwidth}
         \caption{Post-flare segment}
         \includegraphics[scale=0.7, angle=0 ,trim={0 1in 0 0}, clip]{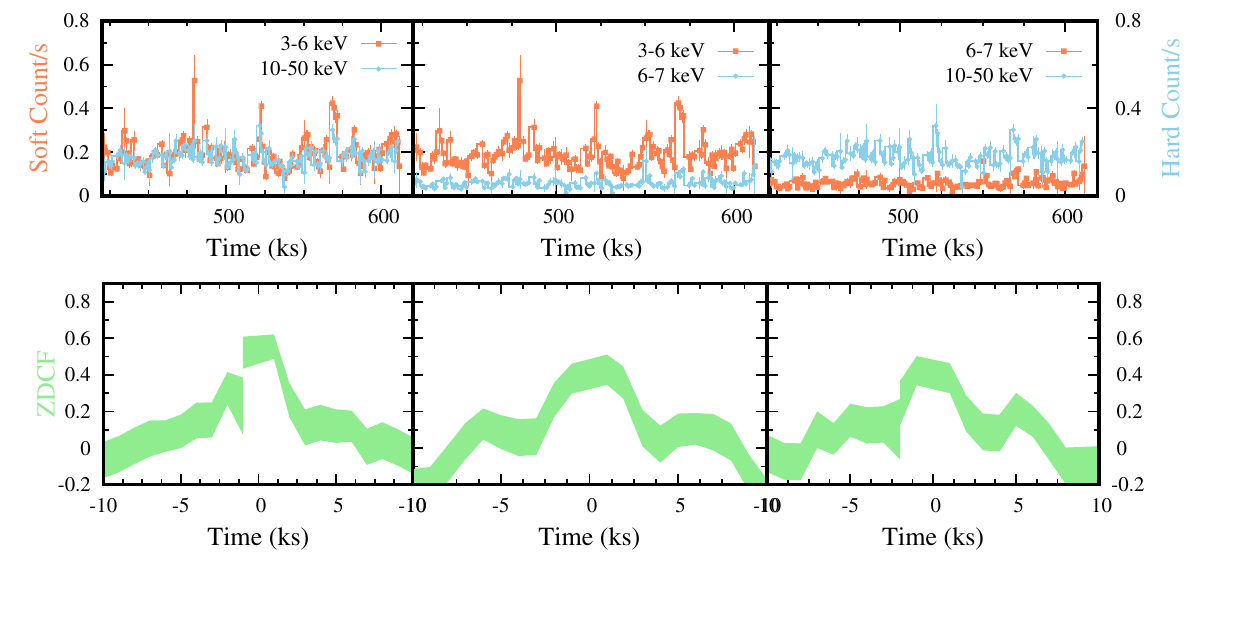} 
         \label{fig:post-flare}
     \end{subfigure}
        \caption{The light curves in $3-6$ keV, $6-7$ keV and $10-50$ keV energy ranges and the correlation function between the light curves in any of the two energy ranges are shown for the pre-flare (a), flare (b) and post-flare (c) segments. For each segment, the light curves are shown in the upper panel, whereas corresponding correlation function between the light curves is shown in the bottom panel. }
        \label{fig:hbb}
\end{figure}

Reflection is one of the explanations for the observed hump in the spectrum above 10 keV. There is another interpretation of the hump put forth by \cite{2021MNRAS.501.5659T}. \cite{2007ApJ...656.1056L} studied the effect of Compton downscattering of the continuum photons in the disc and concluded that the hump-like structure is only formed when the photon index $\Gamma$ of the incident spectrum is less than 2. \cite{2018ApJ...859...89L} claimed that when the photons gain energy up to 511 keV in the vicinity of the black hole, the photon-photon interaction leads to the pair production. The originated positrons then interact with the electrons in a very narrow shell and emit annihilation line photons. To an observer, due to high gravitational redshift ($z \sim 20$), these line photons appear as a broad feature in the spectrum. The argument given in favour of this explanation is a shorter time scale for the hump emission compared to the iron emission line, as these two spectral features are originated in geometrically different regions of the source. The hump emission comes from the closer region to the black hole, while the iron line is emitted from the farther region. 
We estimated the correlation functions among the 3-6 keV, 6-7 keV and 10-50 keV energy bands which are dominated by power-law continuum, iron line and hump or so-called reflection component in pre-flare, flare and post-flare phase, respectively (Figure~\ref{fig:hbb}). We found no correlation between the 6-7 keV and 10-50 keV energy bands in any of the phases, while a moderate correlation was found between the 3-6 keV and 6-7 keV energy bands and 3-6 keV and 10-50 keV energy bands in pre- and post-flaring phases. This indicates that the hump and iron line could be of different origin than the simple reprocessing of the continuum. Therefore, the annihilation line photons scenario could be one of the possible origins of the hump-like spectral feature above 10 keV, often seen in Seyfert galaxies.

\section{Summary \& Conclusions}
\label{sec:summary}
A detailed broadband X-ray spectral and timing analysis of a minor flaring event in NLS1 galaxy NGC~4051 during 4-11 November 2018 has been performed using simultaneous observations with \nustar~ and \xmm~ observatories. While analysing the data, we divided the total exposure into three phases: pre-flare, flare and post-flare and carried out timing and spectral studies during three segments. Following are the main results and conclusions of our analysis:
\begin{itemize}
	\item NGC~4051 showed a rapid variability in the $3-50$ keV energy band of \nustar, where the count rate increased by a factor of $\sim$2.5, followed by a decrease to the pre-flare level within a day. The X-ray variability spectrum of the source showed a trend where variability is higher in soft bands, i.e., soft excess or ionized absorbers. 
	\item From the spectral analysis in the $3-50$ keV range, we found that the spectral slope changes during the observation period. During the flare, the $0.3-50$ keV spectrum was the softest.  
	\item During the X-ray flare, the reflection fraction is the lowest, which is consistent with the increase in coronal height. After the flare, the coronal height reaches its lowest during the observation with the increase in the reflection fraction.
	\item We find that the temperature of the corona increased significantly after the flare (see Figure~\ref{fig:parameter}) from $184_{-28}^{+26}$ to $228_{-52}^{+15}$ keV. This re-heating of the corona is consistent with the hardening of the spectra after the flare subsides. 
	\item In this source, we found signatures of warm absorbers and ultra-fast outflows. 
	\item NGC~4051 hosts a rapidly spinning black hole with spin parameter $a>0.85$ and black hole of mass $\ge 1.32\times 10^5 M_{\odot}$.
	\item NGC~4051 is positioned in the compactness-temperature plane at a region between the electron-electron coupling and the pair runaway line for a slab corona. This implies that the dominant process in the corona is pair production.
	\item Rapid spectral variabilities in NGC~4051 or similar types of NLS1s could be due to the change in the coronal geometry and/or location, which in turn produces small-scale flares. Due to the short time scale of these flares, their origin in the corona is likely to be associated with the ejection of the magnetic field.  
\end{itemize} 

\section*{Acknowledgements}
We thank the anonymous reviewer for the suggestions, which helped us to improve the manuscript. This research work at PRL is funded by the Department of Space, Government of India. AJ acknowledge the support of the grant from the Ministry of Science and Technology of Taiwan with the grand number MOST 110-2811-M-007-500 and  MOST 111-2811-M-007-002. For this research work, the software and online tools provided by the High Energy Astrophysics Science Archive Research Center (HEASARC) online service, maintained by the NASA/GFSC and the High Energy Astrophysics Division of the Smithsonian Astrophysical Observatory have been used. 

\section*{Data Availability}

 We used the archival data from \xmm~ and \nustar~ observatories in this work which is publicly available on HEASARC (\url{https://heasarc.gsfc.nasa.gov/docs/archive.html}).



\bibliographystyle{mnras}
\bibliography{references} 








\bsp	
\label{lastpage}
\end{document}